\documentclass[lettersize, final]{IEEEtran}
\IEEEoverridecommandlockouts
\usepackage{graphicx, amsmath, amssymb, subcaption, url, cite, array, amsthm}
\usepackage[ruled, linesnumbered]{algorithm2e} 
\usepackage{algorithm2e, setspace}
\usepackage[font={small}]{caption}
\usepackage[hidelinks]{hyperref}
\usepackage{tikz}  

\usepackage{geometry}
\newcommand*\circled[1]{\tikz[baseline=(char.base)]{
            \node[shape=circle,draw,inner sep=1pt] (char) {#1};}}

\DeclareMathOperator*{\argmax}{\arg\!\max}

\begin{document}
\title{\huge Enhancing Immersion and Presence in the Metaverse with Over-the-Air Brain-Computer Interface}
\author{Nguyen Quang Hieu, Dinh Thai Hoang, Diep N. Nguyen, Van-Dinh Nguyen, Yong Xiao, and Eryk Dutkiewicz
\thanks{This research was supported by the Australian Research Council under the DECRA project DE210100651, by VinUniversity Seed Grant program, and in part by the National Natural Science Foundation of China under grant 62071193. (\textit{Corresponding author: Van-Dinh Nguyen}.)}
\thanks{Nguyen Quang Hieu, Dinh Thai Hoang, Diep N. Nguyen, and Eryk Dutkiewicz are with the School of Electrical and Data Engineering, University of Technology Sydney, Sydney, NSW 2007, Australia (emails: hieu.nguyen-1@student.uts.edu.au, hoang.dinh@uts.edu.au;  diep.nguyen@uts.edu.au; eryk.dutkiewicz@uts.edu.au).}
\thanks{Van-Dinh Nguyen is with the College of Engineering \& Computer Science and also with the VinUni-Illinois Smart Health Center, VinUniversity, Vinhomes Ocean Park, Hanoi 100000, Vietnam (e-mail: dinh.nv2@vinuni.edu.vn).}
\thanks{Yong Xiao is with the School of Electronic Information and Communications at the Huazhong University of Science and Technology, Wuhan 430074, China, also with the Peng Cheng Laboratory, Shenzhen, Guangdong 518055, China, and also with the Pazhou Laboratory (Huangpu), Guangzhou, Guangdong 510555, China (e-mail: yongxiao@hust.edu.cn).}
}

\maketitle
\begin{abstract}
This article proposes a novel framework that utilizes an over-the-air Brain-Computer Interface (BCI) to learn Metaverse users' expectations. By interpreting users’ brain activities, our framework can optimize physical resources and enhance Quality-of-Experience (QoE) for users.
To achieve this, we leverage a Wireless Edge Server (WES) to process electroencephalography (EEG) signals via uplink wireless channels, thus eliminating the computational burden for Metaverse users' devices.
As a result, the WES can learn human behaviors, adapt system configurations, and allocate radio
resources to tailor personalized user settings. Despite the potential of BCI, the inherent noisy wireless channels and uncertainty of the EEG signals make the related resource allocation and learning problems especially challenging. We formulate the joint learning and resource allocation problem as a mixed integer programming problem. Our solution involves two algorithms: a hybrid learning algorithm and a meta-learning algorithm. The hybrid learning algorithm can effectively find the solution for the formulated problem. Specifically, the meta-learning algorithm can further exploit the neurodiversity of the EEG signals across multiple users, leading to higher classification accuracy.
Extensive simulation results with real-world BCI datasets show the effectiveness of our framework with low latency and high EEG signal classification accuracy.
\end{abstract}

\begin{IEEEkeywords}
Metaverse, brain-computer interface, EEG, edge computing, deep reinforcement learning, meta-learning.
\end{IEEEkeywords}

\section{Introduction}
\subsection{Motivation}
The recent emergence of the Metaverse represents a revolutionary advancement in virtual space, enabling real-time interactions between individuals and objects. 
One of the distinctive features of the Metaverse is that individuals are the central figures, referred to as digital avatars, of all activities and events in the virtual world \cite{xu2022full}. 
In addition to communication and interaction with one another, individuals can also create digital objects in this virtual environment.  However, the real-time representation of human activities in the Metaverse presents a significant challenge for Metaverse development. Most existing solutions rely on an extensive array of sensors to accurately capture human activities in the real world and convert them into the Metaverse \cite{caserman2020survey}. Furthermore, efficient computing and communication solutions are vital for collecting and processing extensive sensing data \cite{xu2022full}. 

To enable Metaverse applications, initial studies have been conducted \cite{meng2022sampling, lam2022human, sun2022augmented, le2022noncontact, fernandes2016combating, cho2014bci, kim2017measurement}. In \cite{meng2022sampling}, the authors proposed a co-design framework for sampling, communication, and prediction to construct virtual objects in the Metaverse. The framework was successfully utilized to reconstruct the trajectories of a virtual robotic arm with low latency, under specific sampling rate and prediction accuracy constraints. The trajectories were captured using a camera as a motion capture system, which collected the movement data from human subjects. Additionally, special markers were placed on the physical object to reconstruct the virtual robotic arm's trajectory. In \cite{lam2022human}, the authors proposed an interactive framework that supports multiple users at a time and a higher degree of freedom for virtual human avatars. To this end, the motion capture system utilized inertial-measurement-unit (IMU) sensors attached to the human participants. The IMU sensors measured the acceleration, velocity, and magnetometer signals of the human joints, which were then used to construct 3D virtual avatars with the iconic gestures of human users.

To further enhance the user experience, the work in \cite{sun2022augmented} proposed a multimodal sensing and feedback platform that could not only detect the finger motion of the users but also send haptic feedback such as vibration and heat to the users from the Metaverse. The haptic feedback was achieved by using triboelectric nanogenerator sensors attached to the user's fingers \cite{le2022noncontact}. The finger motion data were then collected and processed at a cloud server to produce finger motion prediction with machine learning, i.e., Principal component analysis (PCA) and support vector machine (SVM) algorithms. The results showed that the errors in detecting motion and sending haptic feedback were acceptable for collision detection and motion control scenarios.
Other sources of human data such as eye movement, VR headset’s orientation, field-of-view, and heart rate, have shown several benefits in enhancing user expectations \cite{fernandes2016combating, cho2014bci, kim2017measurement}.
For example, VR motion sickness, i.e., a common issue in VR applications in which the user’s brain receives conflicting signals about self-movement between the physical and virtual worlds, can be eliminated by analyzing eye movement, heart rate of the user and adjusting the display settings \cite{fernandes2016combating, cho2014bci}.

As analyzed in the aforementioned works, the utilization of data collected from the Metaverse users can be considered as a potential approach toward a robust and individualized Metaverse world.
However, there are fundamental challenges that need to be addressed to fully realize such an individualized Metaverse world.
One of the main challenges is how we can effectively use such a massive amount of data produced from human activities, given the practical constraints of computing, communications, and energy resources. For example, to make VR headsets less bulky, the battery pack should be smaller, but this might limit the overall energy for local processing and communications. 
This also means fewer antennas and sensors can be mounted, thus limiting the communications and sensing capabilities.
Furthermore, each conventional sensing technique requires a specific type of sensor, e.g., camera \cite{meng2022sampling}, finger sensor \cite{sun2022augmented}, IMU sensor \cite{lam2022human}, heart rate sensor \cite{tidoni2014audio}, and joystick \cite{dasdemir2022brain}, that may hinder the scalability and standardization of the system.
For example, in some Metaverse applications such as real-time virtual fighting, fitness, and dance gaming, heart rate tracking information can help to adjust the tempo and intensity of the music, while body movement tracking information can provide more accurate and precise control over the user's dance movements. However, it is challenging to exploit, combine, and synchronize such diverse information sources from different kinds of sensors, e.g., heart rate sensors and IMU sensors.

Unlike most of the existing works in the literature, we leverage an alternative source of human data from users’ brain activities as a new interactive paradigm in the Metaverse. The brain is the most complex organ in the human body, encoded with the richest information reflecting an individual's cognitive capabilities and perception.
It has been reported that brain signals, such as electroencephalography (EEG), contain underlying information about heart rate, limb movement, intentions, emotions, and eye movement \cite{tidoni2014audio, wu2017evaluation, cheng2020brain}. 
By utilizing brain signals from users, we can potentially develop a human-centric Metaverse that authentically and individually represents humans through their brain activities, while minimizing the need for external sensors attached to the body. Moreover, carrying brain signals over communication channels such as wireless mediums can open up new areas of application, such as brain communications, imagined speech, and semantic communications, enabling users to communicate with one another through their thoughts \cite{lee2022toward}. 
To this end, Brain-Computer Interface (BCI) technology \cite{schalk2004bci2000} provides a neural pathway between the human brain and the outside world through sensors attached to the human scalp, enabling the recording of brain activities and connection to computing units such as remote servers or personal computers.
As such, BCI can be recognized as a facilitator for the enhancement of the human-centric Metaverse, where human activities and thoughts can be reliably mirrored and synergized.

\subsection{Related Works}
\label{subsubsec:realted-works}
Applications of BCI to the Metaverse are still in its infancy.
In \cite{lee2022toward}, the authors proposed a framework for imagined speech communication in home control applications using BCI. The user's EEG signals were extracted and processed with an SVM classifier to detect the user's commands. In another study \cite{dasdemir2022brain}, a gaming platform was developed for teleportation within the Metaverse, where the user's EEG signals were directly translated into commands, eliminating the need for traditional gaming tools like a joystick. The results also showed that participants did not experience motion sickness, as the negativities of the locomotion movements were eliminated by using EEG signals instead of a joystick controller. It is worth noting that both studies in \cite{lee2022toward, dasdemir2022brain} were pseudo-online and supported by a wired connection between a computer and BCI headsets. The local computer was responsible for both processing and providing VR experiences. 
The framework is hence not practical for a scalable Metaverse system due to the mobility and computational limitation of the local computing unit.

The mobility, scalability, and computation limitations of the existing BCI-based Metaverse approaches can be addressed by utilizing wireless BCI headsets connecting to a centralized server/edge computing unit with much more abundant computational power \cite{he2015wireless}. In this way, the remote server and the users can exchange information with each other through the uplink and downlink wireless channels.
For instance, the authors in \cite{kasgari2019human} assumed that delay perception in a wireless network depends on human factors such as age, gender, and demographic, and developed a learning model to predict the perception delay of human users.
The learning model can then allocate radio resources to the users and thus help to minimize the system's delay, constrained by the brain characteristics.
A limitation of \cite{kasgari2019human} is the lack of justification for the brain signals as the authors used delay feedback data from a conventional setting of a mobile network testbed. Furthermore, the brain signals used in \cite{kasgari2019human} were simulated data and oversimplified, making it infeasible to learn human behaviors from their actual brain activities, e.g., EEG signals.
More importantly, the joint problem of predicting human activities and resource allocation has not been addressed in \cite{kasgari2019human}. In reality, these two problems are strongly correlated. For instance, inappropriate resource allocation policies or incorrect prediction of brain activities can severely degrade the user experience, leading to system delays and motion sickness for Metaverse users. The holistic problem becomes even more critical in future wireless systems like 5G advanced and 6G, where massive amounts of user data are exchanged, and ultra-reliable and low-latency requirements must be met.

In our previous work \cite{hieu2022toward}, a BCI-enabled Metaverse system over wireless networks has been developed to jointly optimize resource allocation and brain signal classification.
Unlike \cite{kasgari2019human}, our work in \cite{hieu2022toward} successfully predicted human behaviors from their actual brain signals, i.e., EEG signals.
However, another important issue of incorporating BCI into the Metaverse that has not been concerned in all the above works is the neurodiversity of the brain signals among different users.
Unlike conventional human data, brain signals are highly individual. In other words, the continuous signals such as EEG might be different across users in certain scenarios.
This inevitable neurodiversity has been recently reported in \cite{kang2014bayesian, vezard2015eeg, zhang2017multi}, which have focused on developing robust and scalable classification frameworks for BCI systems.

Given the above, we summarize three major challenges of using BCI for the Metaverse that have not been well addressed in the literature as follows:
\begin{itemize}
\item First, given the noisy wireless environments with uplink communication bottlenecks, how the system can precisely predict human behaviors from their brain signals? This requires not only highly accurate classifiers for brain signals but also effective resource management schemes to alleviate the inherent problems of wireless communications (e.g., fading and interference).
\item Second, given the sophisticated brain signals, how to optimally address the correlation between (i) predicting user behaviors and (ii) minimizing system delay? Such a joint optimization problem is challenging as it must take into account the impact of the noisy wireless environment and low-latency requirements of VR applications, given the uncertainty of the Metaverse users' activities.
\item Third, how to tackle the neurodiversity of the brain signals that impede the scalability and robustness of the Metaverse system involving multiple users?
\end{itemize}

\subsection{Contributions}
To address the aforementioned challenges, this article proposes a novel over-the-air Brain-Computer Interface (BCI) framework to assist in the creation of virtual avatars as human representations in the Metaverse. First, we propose a novel architecture that allows the Metaverse users to interact with the virtual environment while sending the brain signals on uplink wireless channels to Wireless Edge Servers (WES). By utilizing integrated VR-BCI headsets, we can produce a rich source of human data besides conventional sensing techniques for heart rate measurement, eye tracking, or wearable sensors on limbs.
From the collected brain signals, a WES with more computing power, in comparison to the headsets can synthesize and hence orchestra Digital Avatars (DAs) to enhance Metaverse users' QoE. The DAs with up-to-date brain signals can further predict user behaviors such as their physical actions, e.g., moving hands and feet \cite{moioli2021neurosciences}.
With the prediction ability, the DAs can act as intelligent interfaces to support and/or detect movements, decisions, and emotions of the Metaverse users.

To solve the first challenge discussed in Section \ref{subsubsec:realted-works}, we use a real-world BCI dataset \cite{goldberger2000physiobank} to validate the practicability of our system, given the noisy and low-latency requirements of the wireless networks and VR applications, respectively. 
To address the second challenge, i.e., joint prediction and resource allocation, we first formulate the QoE maximization problem subject to practical delay and prediction accuracy requirements. Even neglecting the dynamics/uncertainty of the wireless and users' environment, the resulting problem is a mixed integer programming that is challenging to solve due to the strong correlation between resource allocation and human brain signals prediction problems. To tackle it, we leverage the recent advances of the actor-critic architecture \cite{schulman2017proximal} and the deep convolutional neural network \cite{zhang2021survey} to jointly optimize the system's resources and predict actions of the users based on their brain signals.
To address the third challenge of neurodiversity, we develop a novel meta-learning algorithm that makes the system robust against the increasing number of Metaverse users. Our main contributions are summarized as follows:
\begin{itemize}
\item We introduce a novel over-the-air BCI-enabled Metaverse system.
By collecting and analyzing the brain signals of the users, the system can create intelligent DAs that serve as a neural interface for the Metaverse users. The intelligent DAs can detect user movements such as hands and feet, thus enabling more sophisticated gesture detection applications such as gaming, virtual object control, and virtual teleportation in the Metaverse.
\item We propose an innovative hybrid learning algorithm to address the mixed decision-making (i.e., radio and computing resource allocation optimization) and classification problem (i.e., predicting user behaviors based on the brain signals). Unlike conventional optimization approaches which separately solve these sub-problems, our novel learning framework with deep neural networks directly optimizes the user QoE with practical constraints of the system's delay and prediction accuracy of the brain signals. The proposed hybrid learning algorithm can jointly (i) predict the actions of the users given the brain signals as the inputs and (ii) allocate the radio and computing resources to the system so that the VR delay of the system is minimized. As a result, our approach is more applicable to practical BCI-enabled Metaverse systems with users' dynamic demands as well as strict VR delay requirements. 
\item We develop a highly effective meta-learning algorithm that specifically addresses the impact of neurodiversity in brain signals. By using a novel meta-gradient update process, the proposed meta-learning can better recognize the neurodiversity in the brain signals, compared with the hybrid learning algorithm. Extensive experiments with real-world BCI datasets show that our proposed solution can achieve a prediction accuracy of up to $84\%$. 
More importantly, the simulation results empirically show that our proposed meta-learning algorithm is more robust against neurodiversity, effectively alleviating the prediction accuracy deteriorating when the number of Metaverse users increases.
\item We conduct comprehensive evaluations on the real-world BCI dataset together with a practical VR rendering process at the WES. The results show that our proposed hybrid learning and meta-learning algorithms can provide low delay for VR applications and achieve high classification accuracy for the collected brain signals. More interestingly, our proposed framework can work well with the brain signals distorted by noise when the Metaverse users and the WES communicate with each other over the noisy channels.
\end{itemize}

The rest of our paper is organized as follows. 
In Section \ref{sec:system-model}, we describe our system model in detail.
In Section \ref{sec:learning-algorithms}, we propose two novel algorithms that can effectively address the mixed decision-making and classification problem of our system.
Section \ref{sec:performance-evaluation} presents comprehensive evaluations of our proposed framework with real-world BCI datasets. Conclusions are drawn in Section \ref{sec:conclusion}.

\section{System Model}
\label{sec:system-model}
\begin{figure}[t]
\centering
\includegraphics[width=0.8\linewidth]{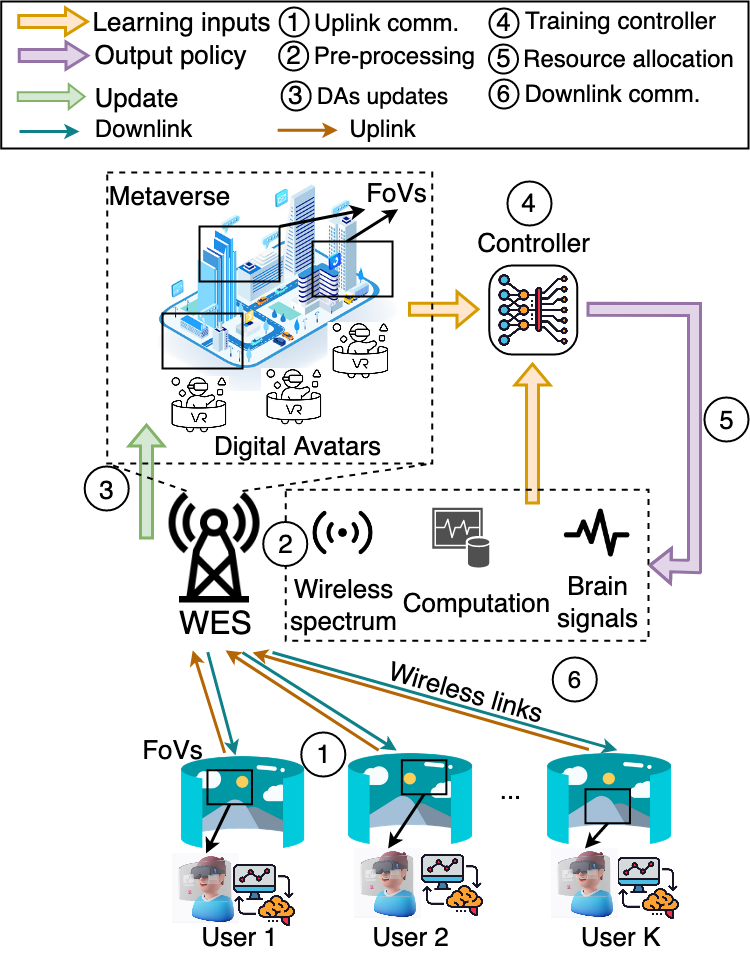}
\caption{Illustration of our proposed over-the-air BCI-enabled Metaverse system. 
A Wireless Edge Server (WES) runs Metaverse applications, supporting VR experiences for $K$ users.
These Metaverse users are equipped with integrated VR-BCI headsets. $K$ Digital Avatars (DAs) are maintained in Metaverse to support real-time recommendations and enhance the user QoE.}
\label{fig:system-model}
\end{figure}

The proposed BCI-enabled Metaverse system model is illustrated in Fig.~\ref{fig:system-model}.
The system consists of (i) a Wireless Edge Server (WES) and (ii) $K$ users equipped with integrated VR-BCI headsets, e.g., Galea headsets \cite{bernal2022galea}.
Each integrated VR-BCI headset can extract brain activities from $J$ channels (i.e., corresponding to $J$ electrodes of the headset) from the user and provide VR services for the user. 
Each user is associated with a Digital Avatar (DA). 
A controller is deployed at the WES to jointly allocate the system's resources and predict user behaviors.
We later describe the deployment of the controller in our proposed algorithms in Section \ref{sec:learning-algorithms}.
The operation of our proposed system includes six main steps as illustrated in Fig.~\ref{fig:system-model}. Details of the system operation are as follows.

\subsection{System Operation}
At each time step $t$, each user sends BCI signals\footnote{We use ``BCI signals" to denote the electroencephalography response signal recorded by an EEG headset. The BCI signals might be slightly different depending on the standard of the BCI system, e.g., 10-10 and 10-20 international systems \cite{jurcak200710}. The details of the BCI system and BCI signals are discussed later.} $\mathbf{e}_k(t)$ to the WES via uplink channels. The BCI signals $\mathbf{e}_k(t) \in \mathbb{R}^{J \times W}$ is a $J \times W$ vector where $J$ is the number of BCI channels, and $W$ is the number of collected BCI signal samples per channel. At the end of the sampling interval of time step $t$, the WES collects a set of BCI signals, i.e., 
\begin{equation}
\mathbf{e}(t) = \{\mathbf{e}_1(t), \mathbf{e}_2(t), \ldots, \mathbf{e}_K(t)\}\in \mathbb{R}^{K \times J \times W}.
\label{eq:bci-signal-vector}
\end{equation}
The above process corresponds to the step \circled{1} in Fig.~\ref{fig:system-model}.
Once the WES obtains BCI signals from the users, the WES pre-processes VR content for $K$ users and monitors the wireless channel state and the internal computing state at the same time (step \circled{2}). 
The pre-processing process can be Field-of-Views (FoVs) rendering that is personalized for each user \cite{fernandes2016combating}. For example, the user may be interested in a particular spatial region in the Metaverse while he/she rarely explores other regions. As a result, pre-processing FoVs for the users can not only save computing resources for the users but also reduce the amount of information transmitted over the wireless links \cite{corbillon2017viewport, hieu2022toward}.
For this, the WES monitors the total computing load of its multi-core CPUs, denoted by $\mathbf{u}(t) \in \mathbb{R}^N$, is defined by:
\begin{equation}
\mathbf{u}(t) = \{u_1(t), u_2(t), \ldots, u_N(t)\},
\label{eq:cpu-load}
\end{equation}
where $N$ is the number of CPUs of the WES and $u_n(t) \in (0, 1)$ is the computing load of the $n$-th CPU.
The WES then analyzes the current state of the system and calculates the best policy, i.e., computing resource allocation for VR pre-processing and radio/power resource allocation for the uplink channels in the next step.

Next, the WES updates the collected BCI signals for $K$ DAs (step \circled{3}). 
We assume that each DA keeps up-to-date BCI signals of a particular user. As such, the creation of intelligent human-like DAs can be easily deployed, maintained, and improved. 
All information, including wireless channels, computing resources, and brain signals, will be used as inputs for the training purpose of the controller (step \circled{4}). 
The controller is a learning model that simultaneously performs two tasks: (i) allocating radio and computing resources of the system and (ii) classifying BCI signals into different actions of the users.
We describe our controller in detail later in Section \ref{sec:learning-algorithms}.
The controller then outputs a policy to increase the QoE of users (step \circled{5}).
Finally, the WES delivers personalized VR services to the users via downlink channels (step \circled{6}). 
The system repeats the above steps in the next time step $t+1$.
As observed from Fig.~\ref{fig:system-model} and the above steps, the neural interface, i.e., integrated VR-BCI headset, can transmit brain signals over wireless channels and allow the WES to deploy VR services that are personalized for the users. For example, by monitoring the EEG signals, the WES can detect VR sickness \cite{lotte2012combining} or emotional changes \cite{cho2014bci} of the users and then adjust the virtual environments' settings accordingly to eliminate such effects.

To evaluate the system's performance, we construct a QoE metric that is a function of (i) the round-trip VR delay of the users and (ii) the accuracy of classifying the BCI signals at the WES. 
The round-trip VR delay is the latency between the time the user requests VR content from the WES (step \circled{1}) and the time the user receives the requested VR content displayed in his/her integrated VR-BCI headset (step \circled{6}). 
The accuracy of classifying BCI signals is obtained by the controller at the WES to predict the actions of the users based on the collected BCI signals.
We select the VR delay and classification accuracy as our main metrics because they have been commonly used to design frameworks that eliminate potential VR sickness or fatigue of the users \cite{chun2016bci, fernandes2016combating, kim2017measurement}. 
Moreover, we consider the classification setting on the BCI signals because if we can successfully predict the actions of the users, it is possible to extend the setting to a general scenario in the Metaverse where intelligent human-like DAs can accurately behave like humans with controlled permissions, e.g., imagined speech communications \cite{lee2022toward}, adaptive VR environment rendering \cite{lotte2012combining}, and anomalous states and error-related behaviors detection \cite{arico2017passive}.
In the sequel, we formally construct the QoE metric by deriving the round-trip VR delay and BCI classifier's accuracy.

\subsection{Round-trip VR delay}
\begin{figure}[t]
    \centering
    \begin{subfigure}[t]{0.49\linewidth}
        \includegraphics[width=1.0\linewidth]{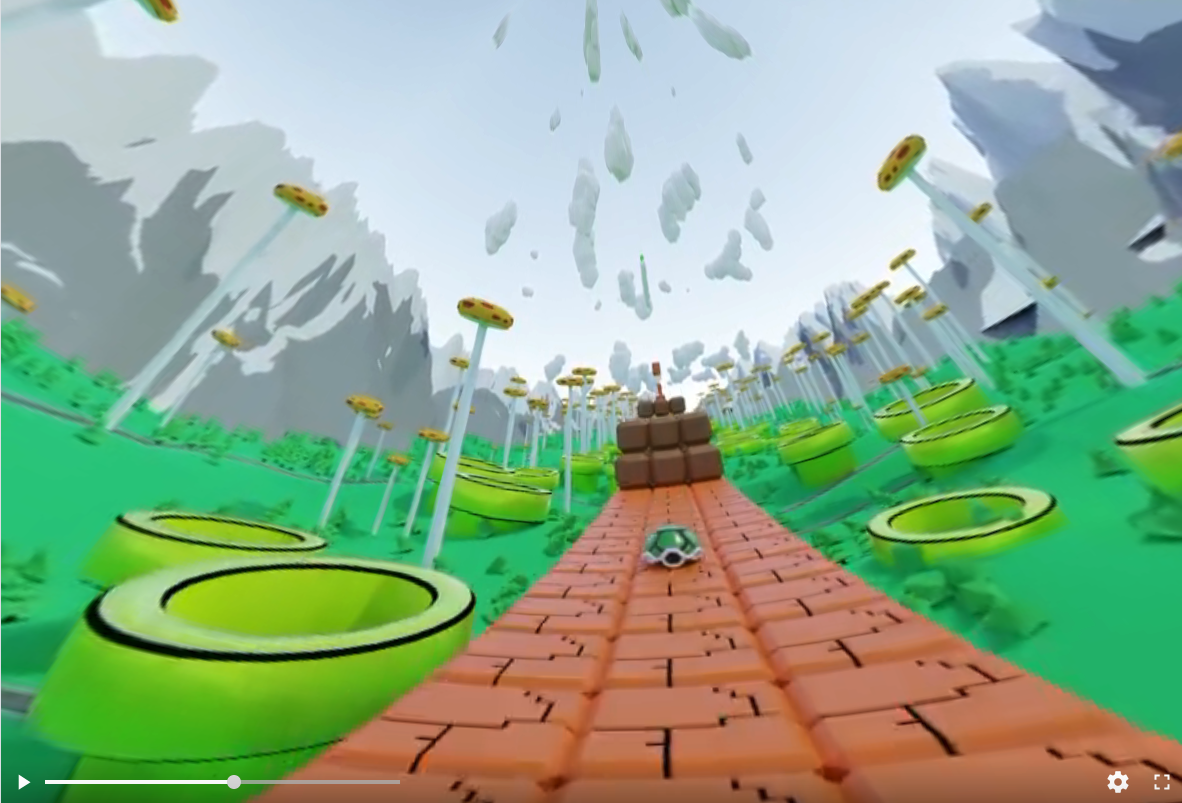}
        \caption{}
    \end{subfigure}
    \begin{subfigure}[t]{0.49\linewidth}
       \includegraphics[width=1.0\linewidth]{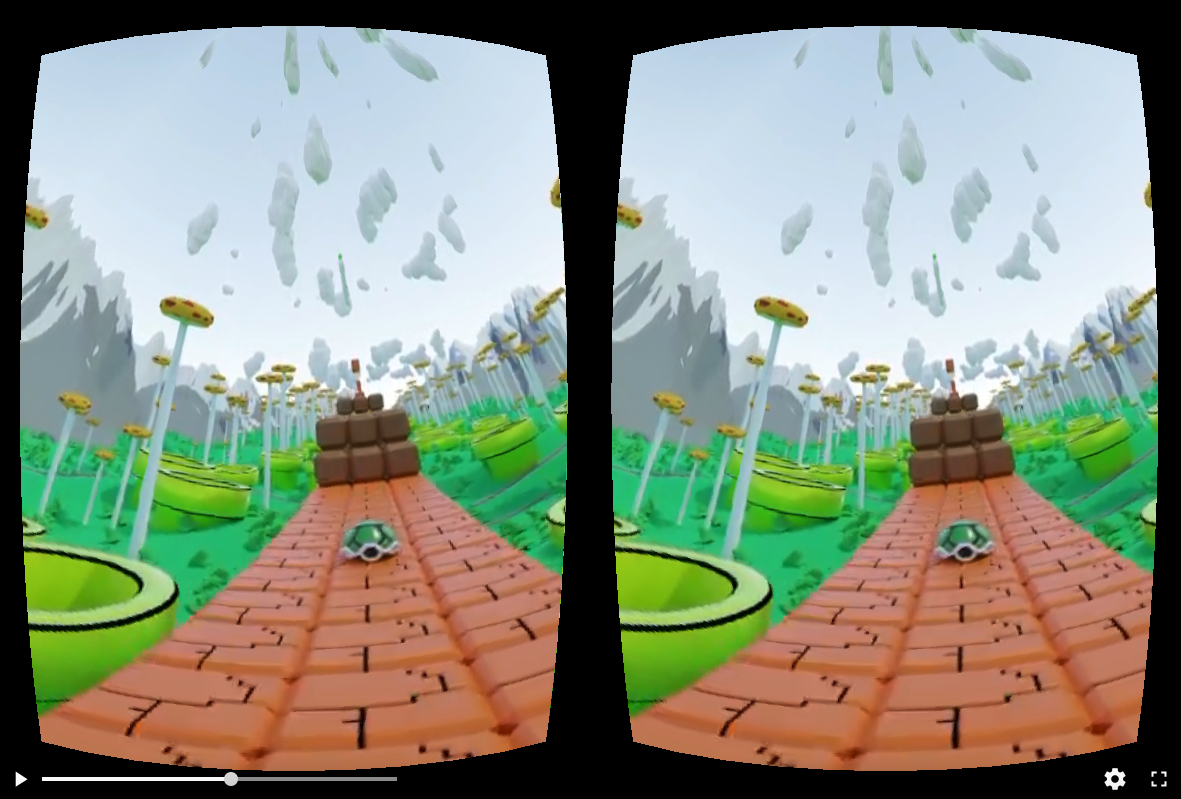}
       \caption{}
    \end{subfigure}
    \begin{subfigure}[t]{0.49\textwidth}
		\centering
		\includegraphics[width=1.0\linewidth]{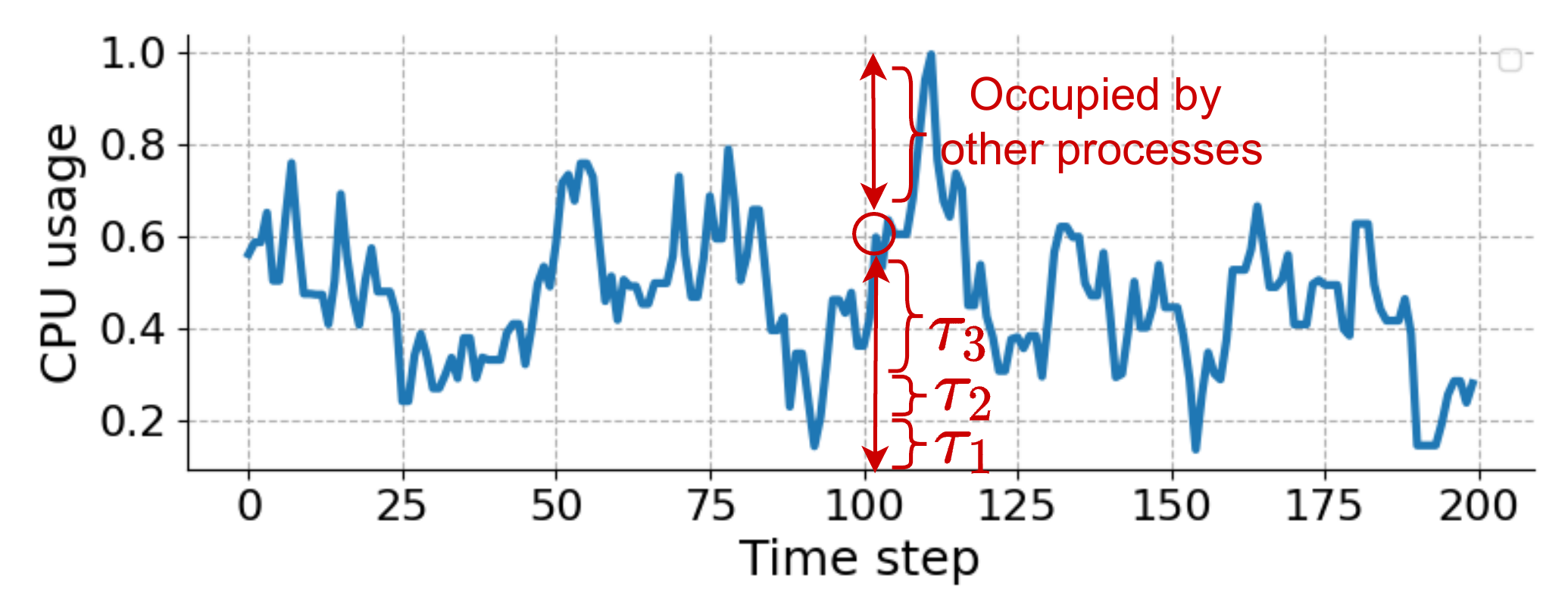}
		\subcaption{}
	\end{subfigure}
    \caption{Illustration of the FoV pre-rendering process at the WES: (a) a selected video frame before rendering and (b) after rendering \protect\footnotemark. (c) illustrates the computing usage of the FoV pre-rendering process. At the time step 100-th, the computing resource is allocated to $K=3$ users, depicted by the red color circles pointing at the CPU usage of 0.6, i.e., $60\%$ CPU.}
    \label{fig:fov-render}
\end{figure}
    \footnotetext{The original video is a panoramic video from Youtube \url{https://www.youtube.com/watch?v=s_hdc_XiXiA}}

We consider that the round-trip VR delay consists of (i) processing latency at the WES, (ii) downlink transmission latency, and (iii) uplink transmission latency. 
For the uplink, we use an orthogonal frequency division multiple access (OFDMA)
technique in which each user occupies a radio resource block and the downlink is broadcast channel \cite{chen2020joint}.
Since most of the computation is shifted to the WES, we assume that the latency at the user headsets is negligible. Accordingly, the round-trip VR delay of user $k$ at time step $t$ is calculated by:

\begin{equation}
D_k(t) = \frac{l_k^U}{r_k^U(t)} + d_k(t) +  \frac{l_k^D}{r_k^D(t)},
\label{eq:system-latency}
\end{equation}
where $l_k^U$ and $l_k^D$ are the length of data packets to be transmitted over the uplink and downlink, respectively; $r_k^U(t)$, $r_k^D(t)$ are the uplink and downlink data rates between the user $k$ and the WES, respectively; and $d_k(t)$ is the processing latency, e.g., pre-rendering the FoVs, at the WES.

The processing delay of the WES depends on the process running in the WES and the CPU capacity of the WES. In our setting, we consider that the WES is running an FoVs rendering process application. We assume that the WES is equipped with a multi-core computing unit having sufficient computing resources for all the users. Our setting is illustrated in Fig.~\ref{fig:fov-render}.
At time step $t$, the WES measures its current available CPU state $\mathbf{u}(t)$.
Let $\tau_k(t) \in (0, 1)$ denote the portion of $u_n(t)$ (i.e., computing load of the $n$-th CPU) that is used for pre-rendering FoV for user $k$-th.
At each time step, the state of the $n$-th CPU is observed by the WES as a normalized value between 0 and 1, which represents the percentage of the current CPU usage. In our later experiments, we empirically obtain the CPU state $n$-th of the WES by running the Vue-VR software \cite{mudin} to pre-render the panoramic video in Fig.~\ref{fig:fov-render}. The CPU usage of the Vue-VR software is then measured as the input training samples for the learning model. In our setting, $K$ users share the same computing resource of CPU $n$-th, so the computing resource allocation variables $\tau_k$ ($k = 1, 2, \ldots, K$) are constrained by $\sum_1^K \tau_k = 1$. Our local server for running the FoV pre-rendering process is a MacBook Air 2020 with 8GB memory and a 2.3 GHz 8-core CPU.

Once $u_n(t)$ and $\tau_k(t)$ are obtained, the pre-processing delay of the WES for rendering FoV for user $k$-th is calculated by:
\begin{equation}
d_k(t) = \frac{1}{\tau_k(t) u_n(t) \upsilon},
\label{eq:processing-delay}
\end{equation}
where $\upsilon$ (Hz) is the CPU capacity, i.e., the total number of computing cycles, of the WES.
The uplink data rate for user $k$ is defined as follows:
\begin{equation}
r_k^U(t) = \sum_{m \in \mathcal{M}} B^U \rho_{k,m}(t) \log_2\Big(1 + \frac{p_{k}(t) h_{k}(t)}{I_m + B^U N_0}\Big),
\label{eq:uplink-rate}
\end{equation}
where $\mathcal{M}$ is the set of radio resource blocks, $p_{k}(t)$ is the transmit power of the user $k$, and $h_{k}$ is the time-varying channel gain between the WES and user $k$. $\rho_{k,m}(t) \in \{0, 1\}$ is the resource block allocation variable. $\rho_{k,m}(t) = 1$ if the resource block $m$ is allocated to user $k$. Otherwise $\rho_{k,m}(t) = 0$. $I_m$ is the multi-user interference from users who are also using the resource block $m$ from nearby WESs. $B^U$ is the bandwidth of each resource block. $N_0$ is the noise power spectral efficiency. 
Note that in our setting, we consider a single cell with one base station, i.e., one WES, and assume that the multi-user interference $I_m$ is negligible or fixed during the simulation episode.

The high interference between multiple users can cause packet errors at the WES.
In our work, we consider the packet error rate experienced by the transmission of BCI signals of user $k$ as \cite{chen2020joint}:
\begin{equation}
\epsilon_k(t) = \sum_{m \in \mathcal{M}} \rho_{k,m} \epsilon_{k, m},
\label{eq:epsilon}
\end{equation}
where $\epsilon_{k,m} = 1 - \text{exp}\Big(- \frac{z \sigma_U^2}{p_k h_k(t)}\Big)$ is the packet error rate over resource block $m$ with $z$ is a waterfall threshold and $\sigma_u^2$ is the noise power of the uplink transmission \cite{xi2011general}. 
Given the observed channel state $h_k(t)$, the inappropriate resource allocations of variables $\rho_{k,m}$ and $p_k$ may cause error transmission with probability $\epsilon_k(t)$. As a result, the received BCI data packet at the WES will be damaged, and the overall system performance will be degraded. From hereafter, we denote the BCI signals containing corrupted packets at the WES as $\mathbf{\hat{e}}(t)$ to differentiate the notation from the error-free BCI signals as defined in (\ref{eq:bci-signal-vector}).

For the downlink channel, the WES can broadcast the VR content to the users. Therefore, the downlink data rate achieved by the WES is calculated by:
\begin{equation}
r_k^D(t) = B^D \log_2 \Big(1 + \frac{P_B h_{k}(t)}{I_D + B^D N_0}\Big),
\label{eq:downlink-rate}
\end{equation}
where $P_B$ is the transmit power of the WES. $I_D$ and $B^D$ are interference and downlink bandwidth, respectively.
As in our considered scenario, the WES can allocate fixed transmission power resources to the users in the downlink channels, and the downlink data packets experience error rates calculated by:
\begin{equation}
\epsilon_k^D(t) = 1 - \exp\Big(-\frac{z \sigma_D^2}{P_B h_k(t)}\Big),
\end{equation} 
where $\sigma_D^2$ is the noise power of the downlink transmission.

\subsection{BCI Classifier}
\label{subsec:bci-classifier}
\begin{figure*}
\centering
\begin{subfigure}{0.35\linewidth}
	\includegraphics[width=1.0\linewidth]{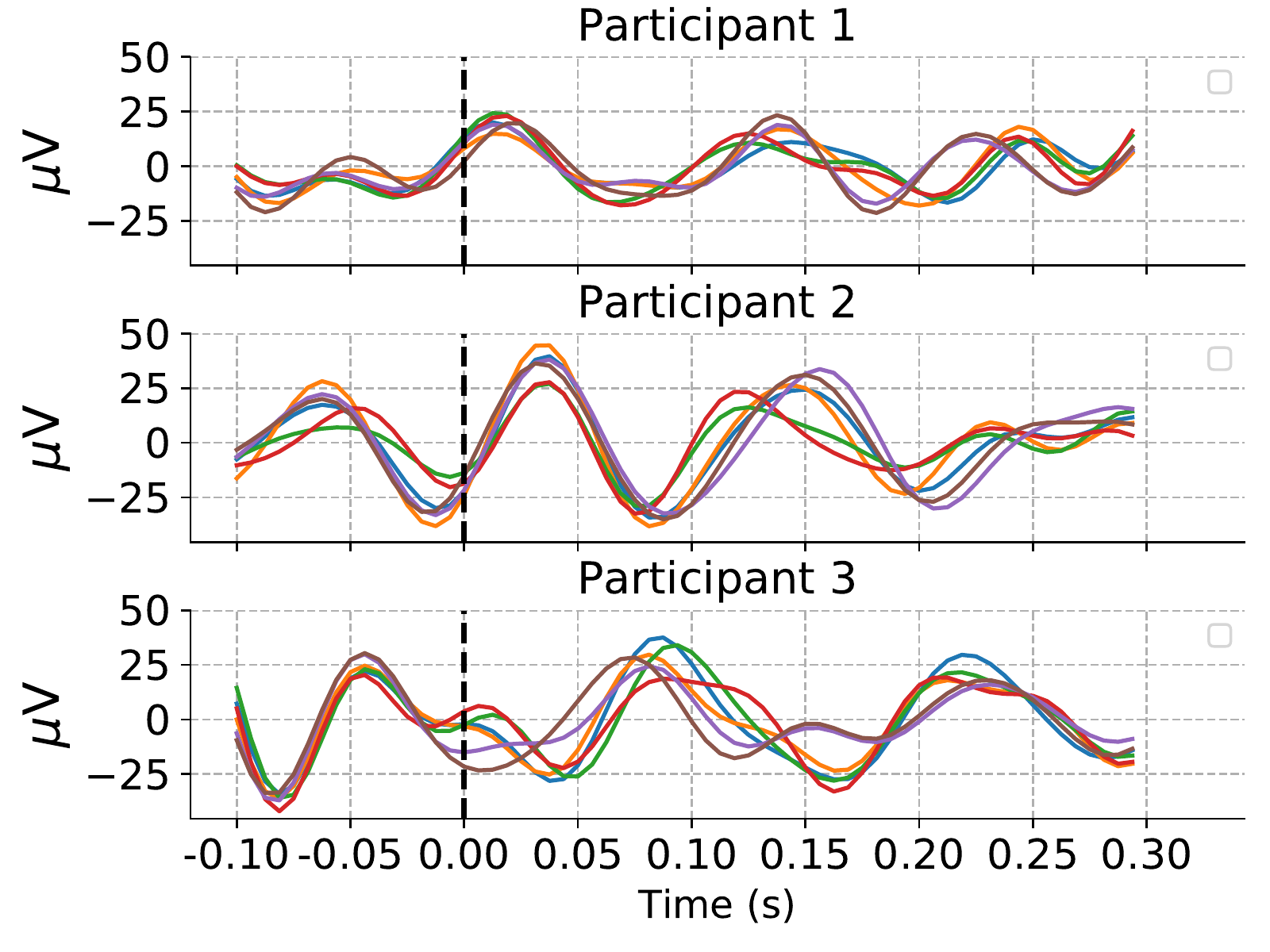}
\end{subfigure}
\begin{subfigure}{0.2\linewidth}
	\includegraphics[width=1.0\linewidth]{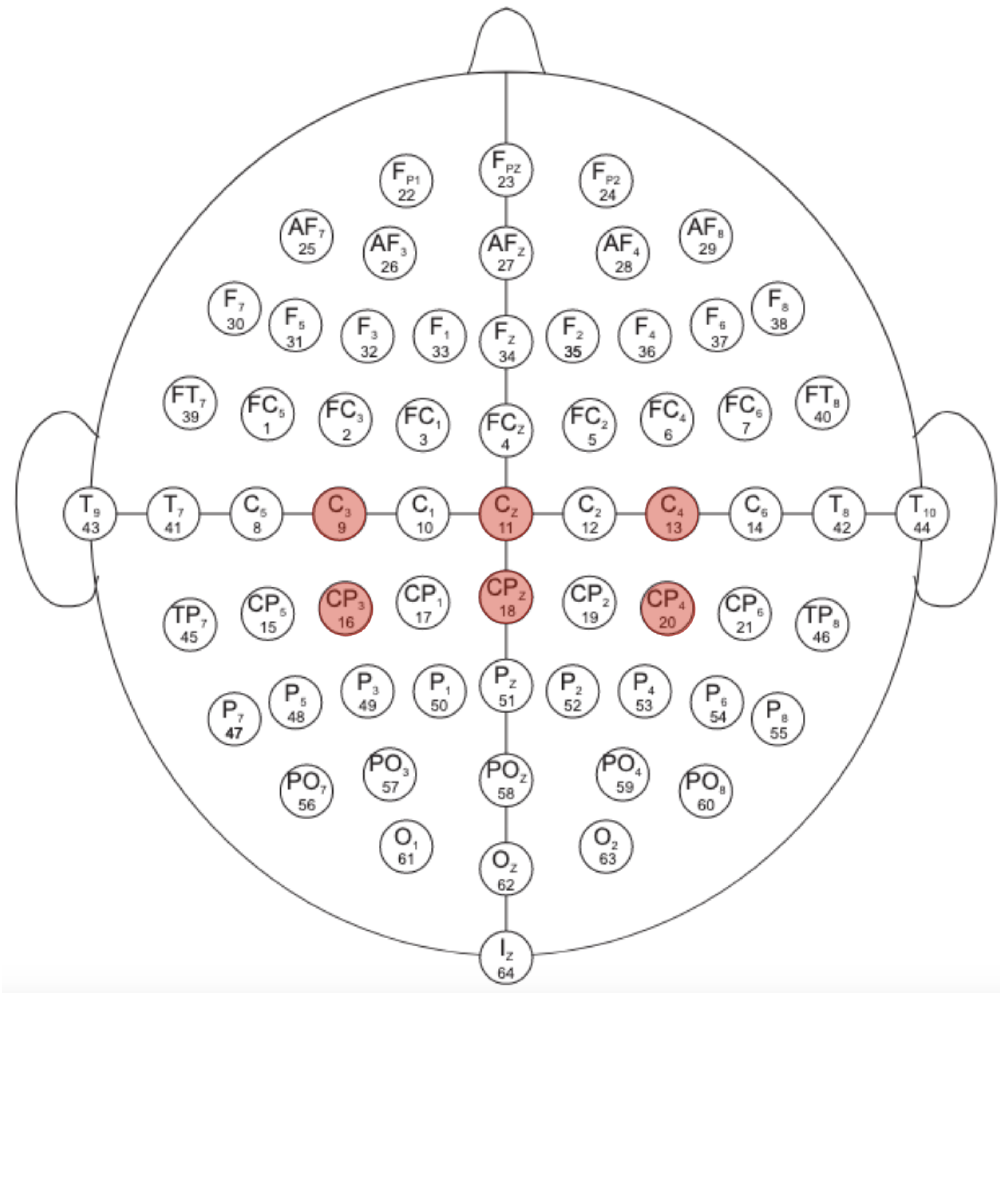}
\end{subfigure}
\caption{Example of EEG signals recorded from three different BCI participants responding to the same experimental condition (left figure). The EEG signals are extracted from the same channels, i.e., C3, CP3, C4, CP4, Cz, and CPz, denoted as red circles on the surface of the scalp in a 10-10 international system (right figure). These channels are responsible for hands and feet movement \cite{morash2008classifying}. The instructions to the participants are placed at the time 0 (marked by the vertical dashed line). The neurodiversity, i.e., subjective differences in the same environment,  reflect the differences in amplitudes and phases of the BCI participants.}
\label{fig:eeg-example}
\end{figure*}

Similar to other works in the literature, we assume that the WES has labels for the input BCI signals \cite{zhang2021survey}.
We consider a BCI classifier at the WES, denoted by $\phi$, to be a binary indicator (0 or 1) if the predicted output, e.g., predicted hands/feet movement, matches the given labels, denoted by $\mathbf{l}(t)$. In particular, $\phi\left(\mathbf{\hat{e}}(t), \mathbf{l}(t)\right) = 1$ if the prediction is correct. Otherwise $\phi\left(\mathbf{\hat{e}}(t), \mathbf{l}(t)\right) = 0$.
The detailed BCI classifier can be derived as follows. The binary indicator function $\phi\big(\mathbf{e}(t), \mathbf{l}(t)\big)$ for the error-free BCI signals $\mathbf{e}(t)$ in (\ref{eq:bci-signal-vector}) is defined by
\begin{equation}
\label{eq:bci-indicator-no-error}
\phi\big(\mathbf{e}(t), \mathbf{l}(t)\big) =
\begin{cases}
1 & \text{if } \argmax \boldsymbol{\varrho} = \mathbf{l}(t), \\
0 & \text{otherwise},
\end{cases}
\end{equation}
where $\boldsymbol{\varrho} = [\varrho_1, \varrho_2, \ldots, \varrho_C]$ is the vector of predicted probabilities for $C$ classes of input BCI signals, i.e., $\mathbf{e}(t)$. The vector $\boldsymbol{\varrho}$ is usually obtained as the output of the learning model, e.g., a deep neural network or a machine learning classifier, given the inputs $\mathbf{e}(t)$ and labels $\mathbf{l}(t)$. The choice of the supervised learning model for the BCI classifier will be discussed further in Sections \ref{sec:learning-algorithms} and \ref{sec:performance-evaluation}. With received BCI data containing error packets, (\ref{eq:bci-indicator-no-error}) can be rewritten as follows:
\begin{equation}
\phi\big(\mathbf{\hat{e}}(t), \mathbf{l}(t)\big) = \big(1 - \epsilon_k^*(t)\big) \phi \big(\mathbf{e}(t), \mathbf{l}(t)\big),
\label{eq:bci-indicator}
\end{equation}
where $\mathbf{\hat{e}}(t)$ is the received BCI data containing error packets due to uplink transmission errors, $\epsilon_k^*(t) = \max_k \epsilon_k(t)$ is the maximum uplink packet error rate (upper-bound error rate) among $K$ users.

The goal of the WES is to minimize the loss of false detections for the predictor $\phi$ given the collected BCI signals and labels.
In our work, we assume that the amount of label data transmitted via uplink channels is negligible, compared with that of the BCI signals. The labels are only scalar values, e.g., 0, 1, and 2, while the corresponding BCI signals are usually sampled at frequencies 150 Hz or 200 Hz \cite{zhang2021survey}.
Formally, we define the loss of the predictor $\phi$ by a cross-entropy loss as follows:
\begin{equation}
\label{eq:cross-entropy-loss}
\begin{split}
L_{\phi}\big(\mathbf{\hat{e}}(t), \mathbf{l}(t)\big) 
& = -\sum_{c=1}^C \phi\big(\mathbf{\hat{e}}(t), \mathbf{l}(t)\big) \log \varrho_c, \\
& = -\sum_{c=1}^C \big(1 - \epsilon_k^*(t)\big) \phi\big(\mathbf{e}(t), \mathbf{l}(t)\big) \log \varrho_c,
\end{split}
\end{equation}
where $C$ is the number of classes of BCI signals, $\varrho_c$ is the predicted probability of actions $c \in \{1, 2, \ldots, C\}$, e.g., moving hands/feet, and $\epsilon_k^*(t) = \max_k{\epsilon_k(t)}$ is the maximum packet error rate (upper-bound error rate) among $K$ users. 

 The cross-entropy loss in (\ref{eq:cross-entropy-loss}) is a summation over $C$ classes of product of a binary indicator $\phi\big(\mathbf{e}(t), \mathbf{l}(t)\big)$, a log-probability $\log \varrho_c$, and a probability $1 - \epsilon_k^*(t)$, which is resulted from the uplink data transmission error. Unlike the conventional cross-entropy function, i.e., $L_{\phi}\big(\mathbf{e}(t), \mathbf{l}(t)\big) = -\sum_{c=1}^C \phi\big(\mathbf{e}(t), \mathbf{l}(t)\big) \log \varrho_c$ \cite{zhang2018generalized}, our modified cross-entropy loss function contains the error probability caused by uplink error transmission, i.e., $\epsilon_k^*(t)$. The reason for the additional parameter $\epsilon_k^*(t)$ in (\ref{eq:cross-entropy-loss}) is that once the received BCI signals contain a corrupted sample from a single user, the BCI classifier's decision will yield an error with probability $1 - \epsilon_k^*(t)$. This modified cross-entropy loss function helps us to capture the effects of the packet transmission errors (i.e., caused by inappropriate resource allocations) on the classification ability of the BCI classifier. In other words, the BCI classifier experiences an additional miss detection outcome, compared to the conventional error-free classification task.

In this work, we consider that BCI signals are EEG signals as the case study. However, the extension beyond EEG, e.g., electrocardiogram (ECG) or electromyogram (EMG), is straightforward.
We collect the EEG signals from a motor imagery experiment \cite{goldberger2000physiobank}. The dataset in \cite{goldberger2000physiobank} contains EEG signals from 109 participants. Each participant produces data samples from 64 EEG channels with the BCI2000 system \cite{schalk2004bci2000}. Details of the experiment can be found in \cite{goldberger2000physiobank}.
In Fig.~\ref{fig:eeg-example}, we illustrate the EEG signals from three different participants responding to the same instruction in the experiment, i.e., moving their hands and feet. 
It can be observed from Fig.~\ref{fig:eeg-example} that the EEG signals of the participant are different in both amplitude and phase. 
This observation expresses the neurodiversity among different users \cite{moioli2021neurosciences}. With the same considered environment, the BCI signals that reflect the user consciousness are different. 
Given the individual BCI signals of multiple users, it is very challenging to obtain accurate predictions on the raw BCI signals, let alone the noisy BCI signals received at the WES after the signals are transmitted over a noisy channel.
In the following, we design an effective QoE model that can capture the impacts of the noisy BCI signals on the system. We then formulate the problem involving (i) a classification problem and (ii) a decision-making problem.

\subsection{QoE Model and QoE Maximization Problem Formulation}
We consider the QoE of user $k$, denoted by $Q_k$, as a combination of (i) round-trip VR delay and (ii) the prediction accuracy for the actions. Therefore, the QoE metric can be expressed as follows:
\begin{multline}
Q_k(\boldsymbol{\rho}, \mathbf{p}, \boldsymbol{\tau}, \phi) = \frac{1}{T} \sum_{t=1}^{T} \Big( \eta_1 \psi \big(D_k(t), D_{\max}\big) + \\ \eta_2 \phi\big(\mathbf{\hat{e}}(t), \mathbf{l}(t)\big) \Big),
\label{eq:qoe-calculation}
\end{multline}
where $\eta_1$ and $\eta_2$ are the positive weighting factors; and $T$ is the time horizon. $\psi(\cdot)$ is also a binary indicator which is defined as follows:
\begin{equation}
\psi \Big(D_k(t), D_{\max}\Big) = 
    \begin{cases}
      1 & \text{if $D_k(t) \leq D_{\max}$,} \\
      0 & \text{otherwise,}
    \end{cases}   
    \label{eq:delay-requirement}
\end{equation}
where $D_{\max}$ is the maximum allowed round-trip VR delay for user $k$.
Recall that the BCI classifier $\phi(\mathbf{\hat{e}}(t), \mathbf{l}(t))$ is a binary indicator that receives value 1 if the classification is correct, as defined in (\ref{eq:bci-indicator}).
The use of binary indicators with positive weighting factors enables the multi-objective QoE model and eliminates the effects of the differences in measurement scales, i.e., time and accuracy. Similar types of multi-objective QoE models have been widely used in the literature \cite{zhang2019drl260, mao2017neural}.

By defining the QoE model as a linear combination of the two binary indicators, we can capture the impacts of (i) wrong classification in (\ref{eq:cross-entropy-loss}) and (ii) exceeding the VR delay requirement in (\ref{eq:delay-requirement}).
For example, an incorrect classification and an exceeded VR delay caused by the WES lead to the QoE value being 0. If both indicators are equal to 1, the QoE value is equal to $\eta_1 + \eta_2$.
By controlling the weighting factors $\eta_1$ and $\eta_2$, one can determine the priorities of such factors, i.e., delay or accuracy, in specific applications.
For example, in applications that require highly accurate classifications of BCI signals such as imagined speech communication \cite{lee2020neural}, the value of $\eta_2$ can be increased. 
Likewise, in delay-sensitive applications, the value of $\eta_1$ can be increased.
The impacts of these weighting factors on the QoE of the users will be further discussed in Section \ref{sec:performance-evaluation}.

In this paper, we aim to maximize the average QoE of users, given the following constraints: (i) power at the WES and user headsets, (ii) wireless channels, and (iii) computational capability of the WES.
Formally, our optimization problem is defined as follows:
\begin{subequations}
\label{eq:min-latency}
\begin{align}
\mathcal{P}_1: \max_{\boldsymbol{\rho}, \mathbf{p}, \boldsymbol{\tau}, \phi} \quad & \frac{1}{K} \sum_{k \in \mathcal{K}} Q_{k}(\boldsymbol{\rho}, \mathbf{p}, \boldsymbol{\tau}, \phi) \\
\textrm{s.t.} \quad & \rho_{k,m}(t) \geq 0, \\
\quad & \sum_{k \in \mathcal{K}} \rho_{k,m}(t) = 1, \\
\quad & 0 \leq p_{k}(t) \leq P_{\max}, \\
\quad & \sum_{k \in \mathcal{K}}\tau_k(t) = 1, \tau_k \geq 0, \\
\quad & \phi\big(\mathbf{\hat{e}}(t), \mathbf{l}(t)\big) \in \{0, 1\},
\end{align}
\label{eq:qoe-maximization}
\end{subequations}
where $P_{\max}$ is the maximum transmission power of the integrated VR-BCI headsets. $\boldsymbol{\rho} = \{\rho_{k, m}(t); \forall k\in \mathcal{K}, \forall m\in \mathcal{M}\}$ is the resource block allocation vector, $\boldsymbol{\tau} = \{\tau_{k}(t); \forall k\in \mathcal{K}\}$ is the computing resource allocation vector, and $\mathbf{p} = \{p_k(t); \forall k \in \mathcal{K}\}$ is the power allocation vector.
In the problem (\ref{eq:qoe-maximization}) above, (\ref{eq:qoe-maximization}b) and (\ref{eq:qoe-maximization}c) are the constraints for radio resource block allocation, (\ref{eq:qoe-maximization}d) is the constraint for the transmit power, (\ref{eq:qoe-maximization}e) are the constraints for the computing resource allocation at the WES, and (\ref{eq:qoe-maximization}f) is the BCI classifier constraint.

Note that the maximization of $Q_k$ in (\ref{eq:qoe-maximization}) results in reducing the round-trip VR delay $D_k(t)$ and the BCI prediction loss $L_{\phi}$. 
Our considered problem involves not only a classification problem, i.e., classification of BCI signals in (\ref{eq:cross-entropy-loss}), but also a decision-making problem, i.e., channel, power, and computing resource allocation problem in (\ref{eq:processing-delay}) and (\ref{eq:uplink-rate}). 
The formulated problem $\mathcal{P}_1$ is a Mixed-Integer Linear Programming (MILP) problem in which the power allocation variables $\mathbf{p}$ are continuous while the resource block allocation and classification decision variables, i.e., $\boldsymbol{\rho}$ and $\phi$ are integer variables. 
Thus, it is challenging to obtain the optimal solution for $\mathcal{P}_1$.
On the one hand, as our problem involves a pattern classification task for EEG signals, the optimization solver for $\mathcal{P}_1$ should be a data-driven approach because the iterative collected data, i.e., EEG signals, may reveal better guidance and unbiased estimation for the optimizer. On the other hand, our problem also consists of a decision-making task in which the physical resources need to be allocated to the appropriate entities. Pattern recognition and decision-making are two classes of tasks that need real-world data to make decisions or predictions. Machine learning is thus the most suitable approach for these tasks because it can handle the data-driven and dynamic nature of the problem better than conventional optimization approaches.

Unlike conventional optimization approaches such as \cite{kasgari2019human} which can only separately solve the sub-problems, i.e., delay perception and resource allocation, and thus cannot enable real-time optimization of user QoE, we propose a novel hybrid learning algorithm to jointly predict the user behaviors based on the BCI signals and allocate radio and computing resources in the system. As a result, our approach is more robust against the dynamic of user demand as well as the uncertainty of the wireless channels.
Later, we propose a highly effective training algorithm based on the idea of meta-learning. With the proposed meta-learning algorithm, we can further enhance the prediction accuracy by dealing with the neurodiversity of the brain signals among multiple users. In the next section, we propose two new algorithms that we refer to as ``Hybrid learner" and ``Meta-leaner".
 
\section{Learning Algorithms For Maximizing QoE}
\label{sec:learning-algorithms}
\begin{figure*}[t]
\centering
\includegraphics[width=0.8\linewidth]{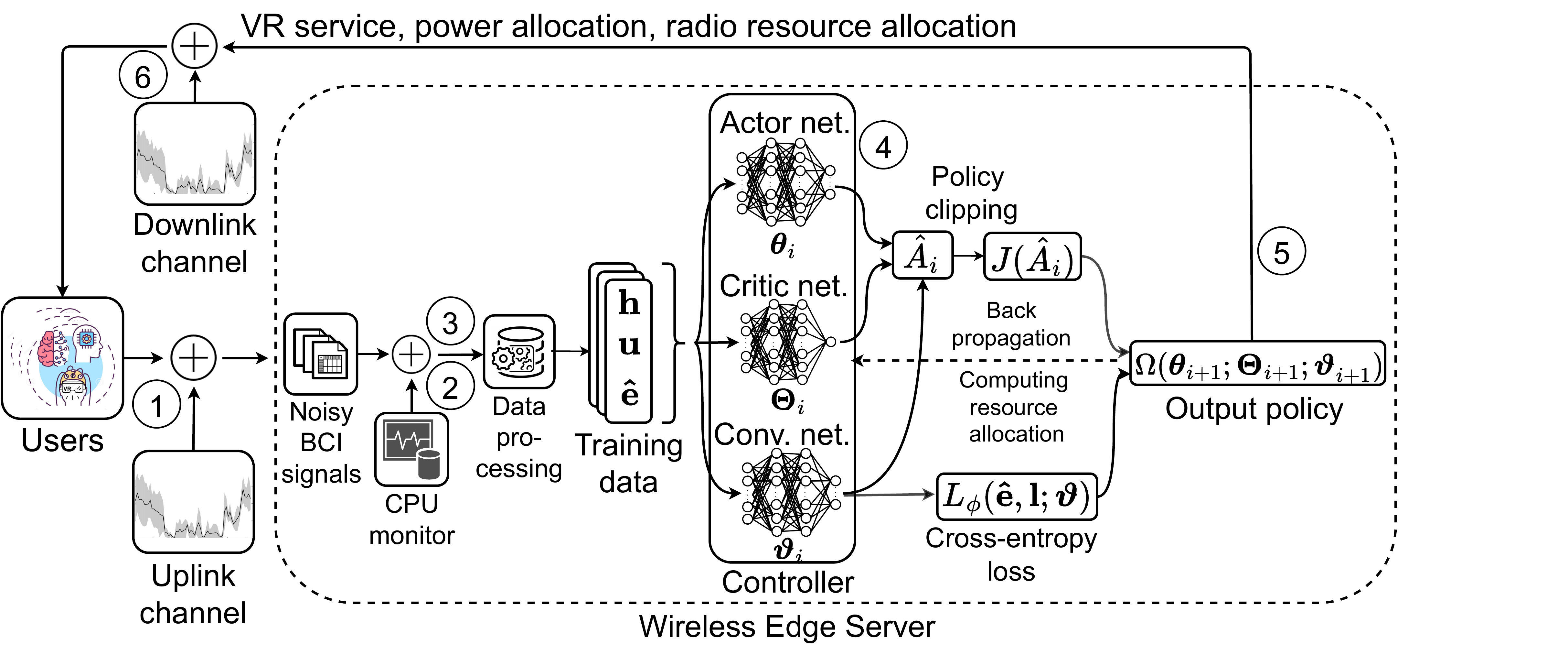}
\caption{Training process for the proposed Hybrid learner at the controller of the WES. The circled numbers denote the corresponding steps as described in Fig.~\ref{fig:system-model} and Section \ref{sec:system-model}.}
\label{fig:hybrid-learner}
\end{figure*}

As we described in Section \ref{sec:system-model}, the controller deployed at the WES is responsible for learning to optimize the system's resources and predict the users' behaviors.
For this purpose, our proposed algorithms, i.e., hybrid learning and meta-learning, in this section can be deployed at the WES just as simply as pre-installing software.
To solve the problem $\mathcal{P}_1$ in (\ref{eq:qoe-maximization}), we propose the Hybrid-leaner to effectively solve the problem.
Next, we propose the Meta-learner as an improvement of the Hybrid learner to solve the problem $\mathcal{P}_2$ in (\ref{eq:meta-optimization}), which will be described later in this section. The problem $\mathcal{P}_2$ is the extended version of $\mathcal{P}_1$ in which the neurodiversity among BCI users is taken into consideration.

\subsection{Hybrid Learning Algorithm}
We first propose a Hybrid learner which is illustrated in Fig.~\ref{fig:hybrid-learner}. 
Our Hybrid learner consists of three deep neural networks that are (i) an actor network, (ii) a critic network, and (iii) a convolutional network.
We note that the convolutional neural network (CNN) can be potentially replaced by other deep learning models such as multilayer perceptron. Among other deep learning models, we adopt a CNN as our BCI classifier because its convolutional layers are more suitable for extracting both spatial and temporal features of the EEG signals, resulting in high accuracy with reasonable computation efficiency.
The inputs for training the deep neural networks are empirical data from the BCI signals, the wireless channel state, and the computing load of the WES. The output of the proposed algorithm is the policy to jointly allocate power for the users' headsets, allocate radio resources for the uplink channels, and predict the actions of the users based on the BCI signals.
Let $\boldsymbol{\theta}$, $\boldsymbol{\Theta}$, and $\boldsymbol{\varphi}$ denote the parameters, i.e., weights and biases, of the actor-network, critic network, and convolutional network, respectively. 
Our proposed training process for the Hybrid learner is illustrated in Algorithm 1. The operation of the algorithm is as follows.

The parameters for deep neural networks are first initialized randomly (line 1 in Algorithm 1).
At each training iteration $i$, the Hybrid learner first collects a set of trajectories $\mathcal{D}_i$ in (\ref{eq:trajectory}) by running current policy $\Omega(\boldsymbol{\theta}_i, \boldsymbol{\Theta}_i, \boldsymbol{\varphi}_i)$ for $O$ time steps. 
The trajectories $\mathcal{D}_i$ contain three main parts that are (i) the observation from the environment, (ii) the action taken of the WES based on the observation from the environment, and (iii) QoE feedback from $K$ users (line 3).
The observation from the environment is a tuple of three states that are channel state $\mathbf{h}$, computing load of the WES $\mathbf{u}$, and BCI signals from users $\mathbf{\hat{e}}$.
The action of the WES is a tuple of four parts that are the radio resource block allocation vector $\boldsymbol{\rho}$, the power allocation vector $\mathbf{p}$, the computing resource allocation vector $\boldsymbol{\tau}$, and the output of the BCI classifier $\phi$.
Based on the collected trajectories, the objective functions for updating the deep neural networks are calculated as follows.
The advantage estimator $\hat{A}_i$ is defined in (\ref{eq:gae-lambda}) \cite{schulman2017proximal} where $\lambda$ is the actor-critic tradeoff parameter and $\delta_o$ is the temporal-difference error which is defined by:
\begin{equation}
 \delta_o = \frac{1}{K} \sum_{k \in \mathcal{K}}Q_{k,o} + \gamma V(\mathbf{h}_{o+1}, \mathbf{u}_{o+1}, \mathbf{\hat{e}}_{o+1}) - V(\mathbf{h}_{o}, \mathbf{u}_{o}, \mathbf{\hat{e}}_{o}),
 \end{equation}
where $\gamma \in (0, 1)$ is the discount factor and $V(\cdot)$ is the value function of the given observation, i.e., the output of the critic network.
Once the advantage estimator is obtained, the decision-making objective can be calculated by $J(\hat{A}_i)$ as defined in (\ref{eq:decision-making-objective}). In the calculation of $J(\hat{A}_i)$, we adopt a policy-clipping technique from \cite{schulman2017proximal}. In particular, the policy clipping function $f_c(\varepsilon, \hat{A}_i)$ is defined by:
\begin{equation}
f_c(\varepsilon, \hat{A}_i)= \begin{cases}(1+\varepsilon) \hat{A}_i, & \text{if } \hat{A}_i \geq 0, \\ (1-\varepsilon) \hat{A}_i, & \text{if } \hat{A}_i < 0.\end{cases}
\end{equation}
With the policy clipping function, the gradient step update is expected not to exceed certain thresholds so that the training is more stable.
Next, the classification loss $L_{\phi}(\mathbf{\hat{e}}, \mathbf{l}; \boldsymbol{\varphi})$ is calculated based on (\ref{eq:cross-entropy-loss}) with convolutional network (line 6).

With all the obtained objective and loss functions, the deep neural networks' parameters are finally updated as follows.
The actor network's parameters are updated in (\ref{eq:actor-net-update}) (line 7) where $\alpha_c$ is the learning step size of the actor network and $\nabla$ is the gradient of the function which can be calculated with stochastic gradient decent/ascent algorithms. In our paper, we use Adam as the optimizer for all the deep neural networks.
The critic network's parameters are updated in (\ref{eq:critic-net-update})   (line 8) where $\alpha_c$ is the learning step size and $L_c(\boldsymbol{\Theta})$ is the critic loss which is defined by:
\begin{equation}
L_c(\boldsymbol{\Theta}) = \Big(V(\mathbf{h}_{o}, \mathbf{u}_{o}, \mathbf{\hat{e}}_{o}) - \frac{1}{K}\sum_{k \in \mathcal{K}}Q_{k,o}\Big)^2.
\end{equation}
Finally, the convolutional network's parameters are updated in (\ref{eq:convo-net-update}) (line 9) where $\alpha_n$ is the learning step size.

As described above, our hybrid learning algorithm addresses the problem $\mathcal{P}_1$ in (\ref{eq:qoe-maximization}) by maximizing the decision-making objective $J(\hat{A}_i)$ in (\ref{eq:decision-making-objective}) while the classification loss $L_{\phi}(\mathbf{\hat{e}}, \mathbf{l}; \boldsymbol{\varphi})$ is minimized in (\ref{eq:convo-net-update}).
With the training process that splits, computes, and backpropagates the losses throughout the deep neural networks, the Hybrid learner can better realize the compound objective $Q_{k}(\boldsymbol{\rho}, \mathbf{p}, \boldsymbol{\tau}, \phi)$ involving distinct learning sub-objectives.
In Section \ref{sec:performance-evaluation}, we show that this design mechanism can significantly enhance the performance of the system, compared with other current state-of-the-art deep reinforcement learning algorithms.

\begin{algorithm}[t]
\caption{Hybrid learning algorithm for maximizing QoE}
 \textbf{Input}: 
 Initialize $\boldsymbol{\theta}_0$, $\boldsymbol{\Theta}_0$ and $\boldsymbol{\varphi}_0$ at random. \\
 \For{\text{i = 0, 1, 2}, $\ldots$}{
  Collect a set of trajectories $\mathcal{D}_i$:
  \begin{multline}
  \mathcal{D}_i = \Big\{\big(\mathbf{h}_o, \mathbf{u}_o, \mathbf{\hat{e}}_o\big), \big(\boldsymbol{\rho}_o, \boldsymbol{p}_o, \boldsymbol{\tau}_o, \phi_o\big), 
  \\ \big(Q_{1,o}, Q_{2,o}, \ldots, Q_{K,o}\big)\Big\}\Big|_{o=1}^{O}.
  \label{eq:trajectory}
  \end{multline} \\
   Compute advantage estimator function $\hat{A}_i$ over $\mathcal{D}_i$:
   \begin{equation}
 \hat{A}_i = \sum_{o=1}^{O}(\gamma \lambda)^o \delta_{o}.
 \label{eq:gae-lambda}
  \end{equation} \\
  Calculate the decision-making objective:
  \begin{equation}
  J(\hat{A}_i) = \min\Big(\frac{\Omega(\boldsymbol{\theta}_i, \boldsymbol{\Theta}_i, \boldsymbol{\varphi}_i)}{\Omega(\boldsymbol{\theta}_{i-1}, \boldsymbol{\Theta}_{i-1}, \boldsymbol{\varphi}_{i-1})} \hat{A}_i, f_c(\varepsilon, \hat{A}_i)\Big).
  \label{eq:decision-making-objective}
  \end{equation} \\
  Calculate $L_{\phi}(\mathbf{\hat{e}}, \mathbf{l}; \boldsymbol{\varphi})$ as defined in (\ref{eq:cross-entropy-loss}). \\
  Update the actor network as follows:
  \begin{equation}
  \boldsymbol{\theta}_{i+1} = \boldsymbol{\theta}_{i} + \alpha_a \nabla J(\hat{A}_i).
  \label{eq:actor-net-update}
  \end{equation} \\
  Update the critic network as follows:
  \begin{equation}
  \boldsymbol{\Theta}_{i+1} = \boldsymbol{\Theta}_{i} - \alpha_c \nabla L_c(\boldsymbol{\Theta}).
  \label{eq:critic-net-update}
  \end{equation} \\
  Update the convolutional network as follows:
  \begin{equation}
  \boldsymbol{\varphi}_{i+1} = \boldsymbol{\varphi}_{i} - \alpha_n \nabla L_{\phi}(\mathbf{\hat{e}}, \mathbf{l}; \boldsymbol{\varphi}).
  \label{eq:convo-net-update}
  \end{equation}
 }
\label{algo:hybrid-learner}
\end{algorithm}

\subsection{Meta-learning Algorithm}
\label{sec:meta-learning}
\begin{algorithm}[t]
\caption{Meta-learning algorithm for better-recognizing neurodiversity}
 \textbf{Input}: 
 Initialize $\boldsymbol{\theta}_0$, $\boldsymbol{\Theta}_0$ and $\boldsymbol{\varphi}_0$ at random. \\
 \For{\text{i = 0, 1, 2}, $\ldots$}{
 
 Collect a set of trajectories $\mathcal{D}_i$ as in (\ref{eq:trajectory}). \\
 Compute advantage estimator function $\hat{A}_i$ over $\mathcal{D}_i$ as in (\ref{eq:gae-lambda}). \\
 Calculate the decision-making objective $J(\hat{A}_i)$ as in (\ref{eq:decision-making-objective}). \\
 \For{\text{k = 1, 2}, $\ldots$, \text{K}}{
 	Compute:
 	\begin{equation}
 	\tilde{\boldsymbol{\varphi}}_k = \boldsymbol{\varphi}_i + g_1 + g_2 + \ldots + g_w.
 	\label{eq:meta-gradient-expand}
 	\end{equation}
 }

 Update the actor-network as in (\ref{eq:actor-net-update}). \\
 Update the critic network as in (\ref{eq:critic-net-update}). \\
 Update the convolutional network as follows:
  \begin{equation}
  \boldsymbol{\varphi}_{i+1} = \boldsymbol{\varphi}_{i} + \alpha_{M} \frac{1}{K} \sum_{k=1}^{K} ( \tilde{\boldsymbol{\varphi}}_k - \boldsymbol{\varphi}_i).
  \label{eq:meta-update}
  \end{equation}
  \\
 }
\label{algo:reptile}
\end{algorithm}

Based on our observation in Fig.~\ref{fig:eeg-example} about the neurodiversity in the brain activities of different BCI users, we are interested in learning a meta-model that jointly optimizes the BCI prediction performance given the BCI signals from different distributions, i.e., phases and amplitudes.
In particular, our optimization problem (\ref{eq:qoe-maximization}) now can be rewritten as:
\begin{subequations}
\label{eq:min-latency}
\begin{align}
\mathcal{P}_2: \max_{\boldsymbol{\rho}, \mathbf{p}, \boldsymbol{\tau}, \phi} \quad & \frac{1}{K} \sum_{k \in \mathcal{K}} Q_{k}(\boldsymbol{\rho}, \mathbf{p}, \boldsymbol{\tau}, \phi) \\
\textrm{s.t.} \quad & (\ref{eq:qoe-maximization}\text{b})-(\ref{eq:qoe-maximization}\text{e}), \\
\quad & \phi_k\big(\mathbf{\hat{e}}_k(t), \mathbf{l}_k(t)\big) \in \{0, 1\},
\end{align}
\label{eq:meta-optimization}
\end{subequations}
where $\phi_k\big(\mathbf{\hat{e}}_k(t), \mathbf{l}_k(t)\big)$ is now the BCI classifier of the user $k$-th, given the noisy BCI signals $\mathbf{\hat{e}}_k(t)$ and labels $\mathbf{l}_k(t)$ sampled from the user $k$-th.
As observed from the problems $\mathcal{P}_1$ and $\mathcal{P}_2$, the difference between the two problems is the constraint $(\ref{eq:qoe-maximization}f)$ and $(\ref{eq:meta-optimization}c)$.
In $\mathcal{P}_1$, we consider that the BCI signals share the same distribution and we ignore the fact that the BCI signals might be significantly different in terms of amplitudes and phases, i.e., neurodiversity \cite{kang2014bayesian, vezard2015eeg, zhang2017multi}. As a result, the trained BCI classifier may obtain a higher performance with the BCI signals from a BCI user than that from another user.
With the formulated problem in $\mathcal{P}_2$, we aim to improve the performance of the BCI classifier, regardless of the neurodiversity when the number of BCI users increases. 

As observed from problem $\mathcal{P}_2$ in (\ref{eq:meta-optimization}), a naive solution can be implementing $K$ Hybrid learners for $K$ BCI classifiers to deal with the diverse distributions problem. In this way, we have to train and maintain multiple learning models for the single purpose of improving the BCI prediction accuracy. However, this implementation may require a switch operator that selects the optimal model given a random or unknown input of BCI signals. This assumption about the switch operator might not be feasible in practice \cite[Ch. 7]{goodfellow2016deep}. Moreover, maintaining multiple learning models can cause extra costs of training and maintenance, thus yielding negative impacts on the servers/systems. 
In our work, we aim to train a single Meta-learner that can achieve high prediction accuracy on the BCI signals with unknown and diverse distributions. 

The proposed meta-learning algorithm is proposed in Algorithm 2. In particular, the detailed training process is as follows.
First, the parameters of the three deep neural networks are initialized at random (line 1 of Algorithm 2).
Similar to Algorithm 1, the Meta-leaner first computes the decision-making objective $J(\hat{A}_i)$ (lines 3, 4, and 5 of Algorithm 2).
Next, the Meta-learner computes $w$-step meta-gradients, denoted by $\tilde{\boldsymbol{\varphi}}_k$, expanded from the current parameter $\boldsymbol{\varphi}_i$ as in (\ref{eq:meta-gradient-expand}) (line 7).
In (\ref{eq:meta-gradient-expand}), $g_1, g_2, \ldots, g_w$ are the gradients computed by SGD (Stochastic Gradient Decent) over $w$ mini-batches of the inputs, i.e., $\mathbf{\hat{e}}_k(t)$.
In other words, the inputs $\mathbf{\hat{e}}_k(t)$ are divided into $w$ mini-batches to compute $w$ gradients $g_1, g_2, \ldots, g_w$.
After that, the actor-network and critic network are updated based on equations (\ref{eq:actor-net-update}) and (\ref{eq:critic-net-update}), respectively (lines 9 and 10).
Finally, the convolutional network is updated with (\ref{eq:meta-update}), where $\alpha_{M}$ is the meta-learning stepsize (line 11).
The calculation of meta-gradients and meta-updates in equations (\ref{eq:meta-gradient-expand}) and (\ref{eq:meta-update}) over $K$ users helps to learn a policy that is equivalent to a distilled policy of $K$ separate Hybrid learners.
During the training phase, each training sample from user $k$ will be included in the gradient computation of equations (\ref{eq:meta-gradient-expand}) and (\ref{eq:meta-update}), which helps the Meta-learner to obtain a refined gradient across all $K$ tasks.
During the testing phase, the Meta-learner receives random EEG signal samples from the users without awareness of which EEG samples belong to which users.

We develop our meta-learning algorithm based on a scalable first-order meta-learning algorithm, named Reptile~\cite{nichol2018first}. 
Similar to the Reptile, our algorithm belongs to the gradient-based meta-learning family in which the gradient computation requires extra steps as described in equations (\ref{eq:meta-gradient-expand}) and (\ref{eq:meta-update}). In return for extra gradient computations, the gradient-based meta-learning algorithm is suitable for any type of deep learning model, i.e., model-agnostic \cite{finn2017model}.
Note that unlike the original Reptile algorithm which is only applicable to the classification problem, our Meta-learner can deal with the joint decision-making and classification problem, thanks to the hybrid design.

In the next section, we empirically show that our proposed hybrid learning algorithm can outperform current state-of-the-art reinforcement learning algorithms in dealing with the decision-making problem. Furthermore, we show that our proposed meta-learning algorithm significantly improves the BCI prediction performance, compared with our proposed hybrid learning and other supervised learning algorithms.
Thanks to the novel meta-learning process, our Meta-learner can understand the neurodiversity at a deeper level than the hybrid learning algorithm, resulting in higher and more robust classification performance under different settings.
Note that the proposed meta-learning algorithm only requires additional gradient computations and the Meta-learner can be deployed at the controller of the WES similar to the Hybrid learner in Fig.~\ref{fig:hybrid-learner}.

\section{Performance Evaluation with BCI Datasets}
\label{sec:performance-evaluation}
\subsection{Data Preprocessing}
\begin{table*}[t]
\centering
\begin{tabular}{c|l|l|c|l|l}
\hline 
\textbf{Notation}  & \textbf{Communication } & \textbf{Default} & \textbf{Notation}  & \textbf{Algorithmic} & \textbf{Default} \\ 
 & \textbf{Parameters} & \textbf{Settings} &  & \textbf{Parameters} & \textbf{Settings} \\ \hline
$M$ & Number of radio & $10$ resource & $O$ & Trajectories' length & 100 steps\\ 
& resource blocks & blocks & & & \\
\hline$K$ & Number of users & $[3: 7]$ users & $T$ & Total training time steps & $3 \times 10^5$ steps\\
\hline$P_B$ & Power of the WES & $1$ W~\cite{chen2020joint} & $\lambda$ & Advantage estimator's  & 0.99 \cite{schulman2017proximal} \\
& & & & parameter & \\
\hline$P_{\max}$ & Power of the user headset & [-20: 20] dBm & $\gamma$ & Advantage estimator's  & 0.99 \cite{schulman2017proximal}\\
 & & & & parameters &  \\
\hline$B^U$ & Uplink bandwidth & $1$ MHz~\cite{chen2020joint} & $\varepsilon$ & Clip ratio & 0.2 \cite{schulman2017proximal} \\
\hline$B^D$ & Downlink bandwidth & $20$ MHz~\cite{chen2020joint} & $\alpha_a$ & Actor network's  & $5 \times 10^{-5}$ \\
& & & & learning rate & \\
\hline$N_0$ & Noise & $-174$ dBm & $\alpha_c$ & Critic network's & $5 \times 10^{-4}$ \\
& & & & learning rate &\\
\hline$I_m$ & Interference & $-10$ dBm & $\alpha_n$ & Convo. network's & $2 \times 10^{-3}$ \\
$I_D$ & & & & learning rate &\\
\hline$D_{\max}$ & Maximum round-trip delay & $10$ milliseconds & $\alpha_{M}$ & Meta-learning rate & 1.0 \cite{nichol2018first}\\

\hline 
\textbf{Notation} & \textbf{Computation} & \textbf{Default} & \textbf{Notation} & \textbf{BCI Parameters} & \textbf{Default} \\ 
 & \textbf{Parameters} & \textbf{Settings} &  & & \textbf{Settings} \\ \hline
$\upsilon$ & Computation capacity of & $[2.3 \times 10^{-6}: 2.3]$ & & BCI dataset & \cite{goldberger2000physiobank} \\
            & the WES        & GHz & & & \\
\hline & Video quality for the VR   & $1280 \times 702$ & $J$ & Number of BCI channels & 64 channels \\
            &  processing & pixels & & &\\
\hline 
& & & $C$ & Number of class labels & 4 classes \\ \hline
& & &  & Number of BCI users & [3:7] users \\ \hline
& & &  & Sampling rate & 160 Hz \\ \hline
\end{tabular}
\caption{Parameter settings.}
\label{tab:simulation-settings}
\end{table*}

We conduct extensive simulations to evaluate the system performance as follows.
For the BCI classification problem, we use a public dataset from~\cite{goldberger2000physiobank}. The dataset contains the experiment results of 109 participants. Each participant is instructed to do an action per experimental run. The actions are opening/closing eyes, fists, and feet. 
On each experimental run, the EEG signals are obtained through 64 EEG channels with the BCI2000 system~\cite{schalk2004bci2000}. The sampling rate is 160 Hz. In our setting, we consider four different actions, i.e., $C=4$, that are open eyes, close eyes, close fist, and move feet.
In the default setting, we consider BCI signals from three users as illustrated in Fig.~\ref{fig:eeg-example}. 
We adopt a data processing pipeline from \cite{zhanggithub}. In particular, the data processing is as follows.
The collected BCI signals of each user have 255,680 data samples. 
Because EEG signals are temporal data, we split the data stream into different segments and iteratively input the segments into the deep neural networks.
Each segment contains 16 EEG samples, which is equivalent to 0.1 second as the sampling frequency is at 160 Hz. 
The overlapping rate between two adjacent segments is set at 50\%.
After segmentation, the data samples are then normalized with the z-score normalization technique \cite{zhanggithub}.
Finally, the data samples are split into a training set and a test set with a ratio is 80:20, respectively.

Note that similar to other BCI research works in the literature, we consider a classification setting with a discrete number of actions \cite{zhang2021survey}. For capturing full human body movement with a high degree of freedom (DoF), current BCI technologies are not yet ready because of the complexity of BCI signals. 
In this work, we only focus on enabling the potential of BCI for the Metaverse without focusing on realistic avatar/human motion capture techniques.
Human motion capture is another topic that is outside the scope of our work. Further details of the data collection process of a BCI experiment, i.e., a motor imagery decoding experiment, can be found in Appendix \ref{sec:appendix}.

The Hybrid learner's architecture is set as follows.
The actor network and critic network are multilayer perceptron networks (MLPs) that consist of one input layer, two hidden layers, and one output layer.
In our default setting, the number of users is set at $K=3$.
In this case, the number of input neurons at the input layer is 6, which is equivalent to the total sizes of the channel state and the CPUs, i.e., $\{h_1(t), h_2(t), h_3(t), u_1(t), u_2(t), u_3(t)\}$.
Note that we only measure the 3 most available CPUs of the WES to feed into the deep neural networks for simplicity.
The number of the output neurons at the output layer is 9, which corresponds to the number of radio resource block allocation variable, the power allocation variable, and the computing resource allocation variable, i.e., $\{\rho_1(t), \rho_2(t), \rho_3(t), p_1(t), p_2(t), p_3(t), \tau_1(t), \tau_2(t), \tau_3(t)\}$.
For the CNN, we use each EEG channel as feature input for the CNN of the Hybrid learner. Thus, we have 64 input features and 4 class labels to train with the CNN.
We use Adam to optimize the parameters of the deep neural networks.
The Meta-learner reuses the same architecture as the Hybrid learner. 
The difference is computing the meta-gradients and meta-updates in equations (\ref{eq:meta-gradient-expand}) and (\ref{eq:meta-update}), respectively.
For this, an additional SGD algorithm is used to calculate $w$ gradient steps with respect to $w$ mini-batches of the input data, i.e., equation (\ref{eq:meta-gradient-expand}). 

For the decision-making problem, i.e., radio and computing resource allocation, we conduct an experiment as illustrated in Fig.~\ref{fig:fov-render} to measure the processing latency at the WES. For the uplink and downlink latency, we use the Rayleigh fading to simulate the dynamics of the time-varying wireless channel. The number of radio resource blocks is set to $M=10$. The number of Metaverse users is varied from  $K=3$ to $K=7$ users. 
The details of our parameter settings are shown in Table.~\ref{tab:simulation-settings}.
In comparison with our proposed algorithms, we introduce the following baselines.

\textbf{Proximal Policy Optimization (PPO)} \cite{schulman2017proximal}: PPO is a state-of-the-art reinforcement learning algorithm for decision-making problems with continuous action values. Our Hybrid learner also adopts the actor-critic architecture and policy clipping techniques from PPO to achieve robust performance. In particular, the PPO baseline contains an actor network and a critic network. We directly use this architecture to learn the QoE defined in (\ref{eq:qoe-calculation}). By maximizing the average QoE, the PPO baseline is expected to reduce the loss $L_{\phi}$ and the round-trip VR delay $D_k$.

\textbf{Vanilla Policy Gradient (VPG)} \cite{sutton2018reinforcement}: VPG is a classic policy gradient algorithm for decision-making problems with continuous action values. The VPG baseline also uses the actor-critic architecture. However, the VPG algorithm does not have the embedded advantage function and the policy clipping technique. 

\textbf{Support-Vector Machine (SVM)} \cite{cortes1995support}: SVM is a classic supervised learning algorithm and is a robust benchmark for classification problems. In our simulations, we use SVM as a baseline to evaluate the classification accuracy of our proposed algorithms that are developed on deep neural networks. For a fair comparison, we consider the following setting to give SVM advantages compared with our proposed algorithms. First, we replace the convolutional network in our Hybrid learner with the SVM and we keep the actor-network and critic network similar to those of the Hybrid learner. As a result, the SVM-based Hybrid learner can still deal with both decision-making and classification problems. Second, we train the SVM with training data that are collectively fed into the input of the SVM. In other words, all the training data is stored and reused at the WES. We observe that this training method can significantly boost the performance of the SVM. Otherwise, if we apply the same training method as our proposed algorithms, i.e., the training data at each time step is removed after feeding into the deep neural networks, the performance of the SVM is significantly decreased.

\subsection{Experiment Results}
\subsubsection{Convergence of the algorithms on training set}
\begin{figure*}[t]
	\centering
	\begin{subfigure}[b]{0.31\linewidth}
		\centering
		\includegraphics[width=1.0\linewidth]{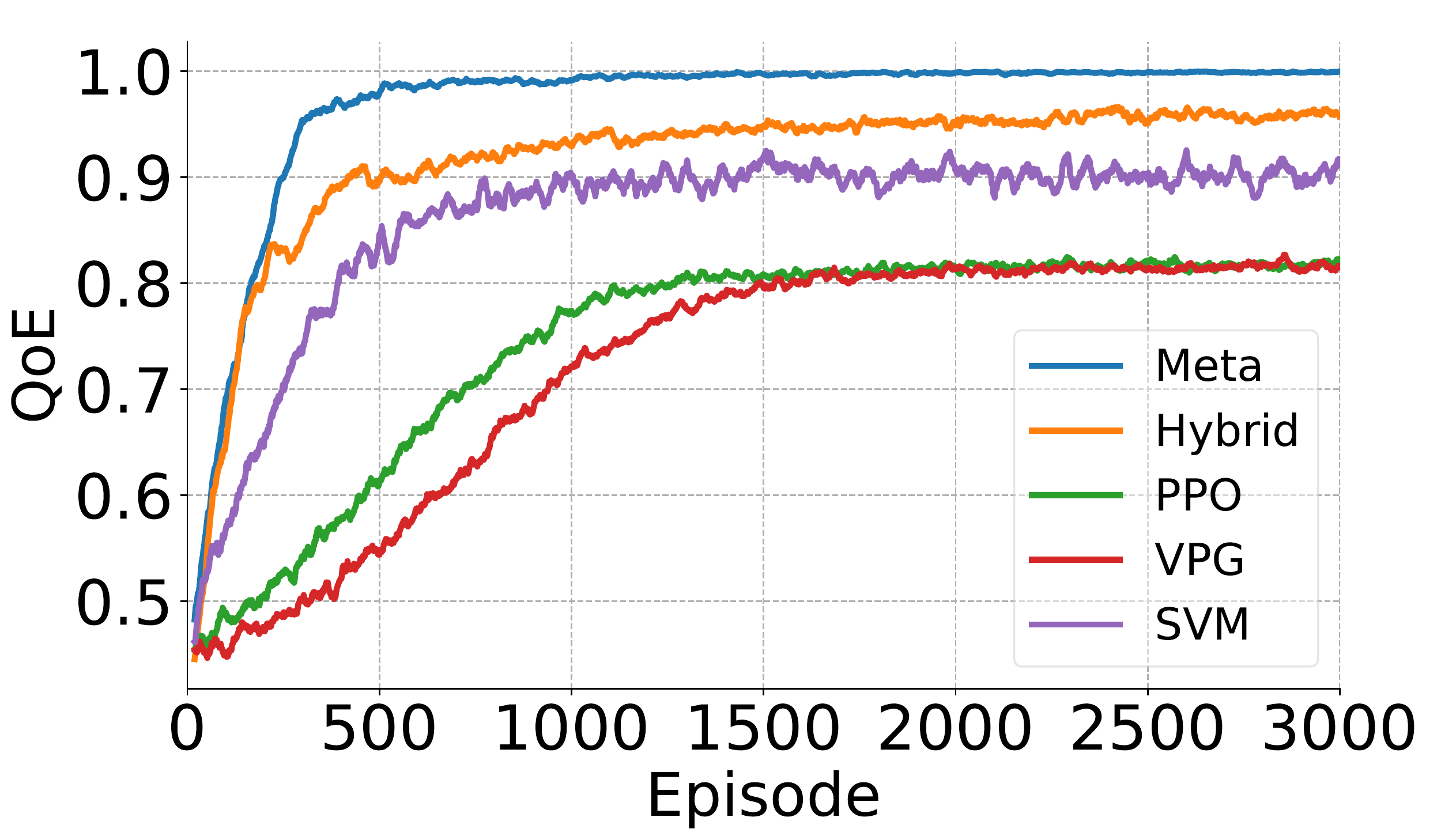}
		\caption{}
	\end{subfigure}%
	~ 
	\begin{subfigure}[b]{0.31\linewidth}
		\centering
		\includegraphics[width=1.0\linewidth]{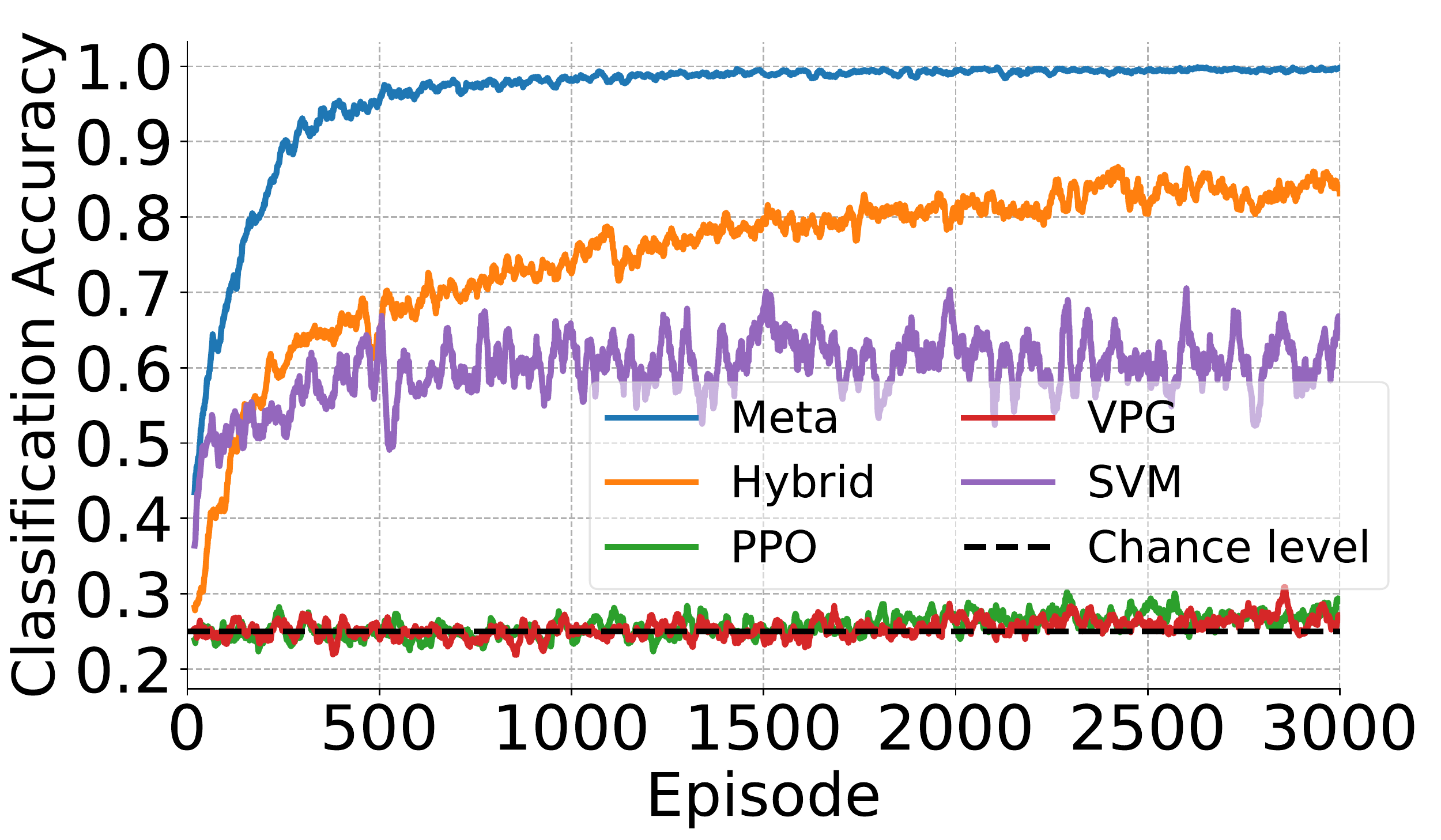}
		\caption{}
	\end{subfigure}%
	~
	\begin{subfigure}[b]{0.31\linewidth}
		\centering
		\includegraphics[width=1.0\linewidth]{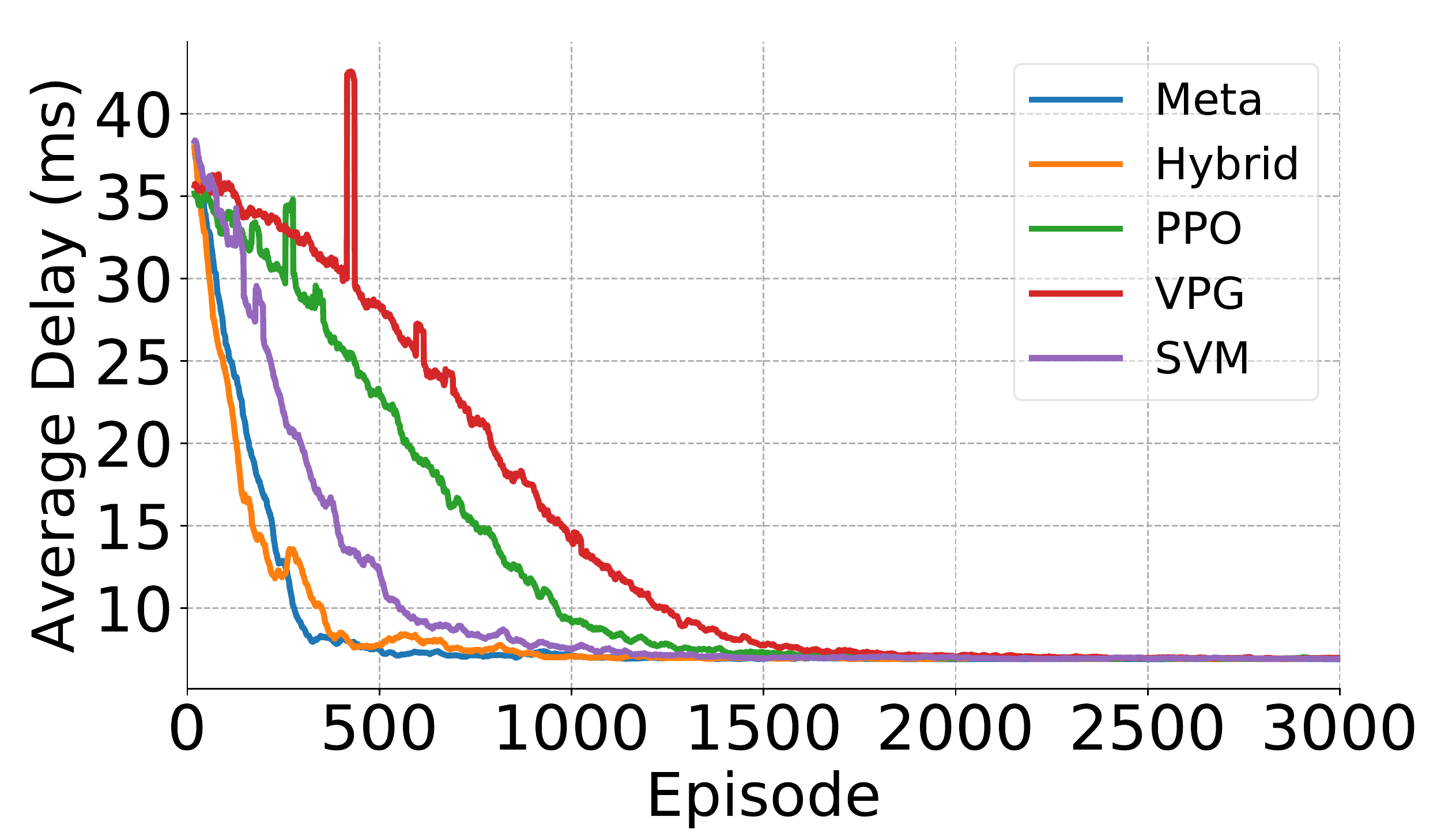}
		\caption{}
	\end{subfigure}
	\caption{(a) Normalized QoE, (b) classification accuracy, and (c) average round-trip VR delay values.}
	\label{fig:training-results}
\end{figure*}

\begin{figure*}[t]
	\centering
	\begin{subfigure}[b]{0.22\linewidth}
		\centering
		\includegraphics[width=1.1\linewidth]{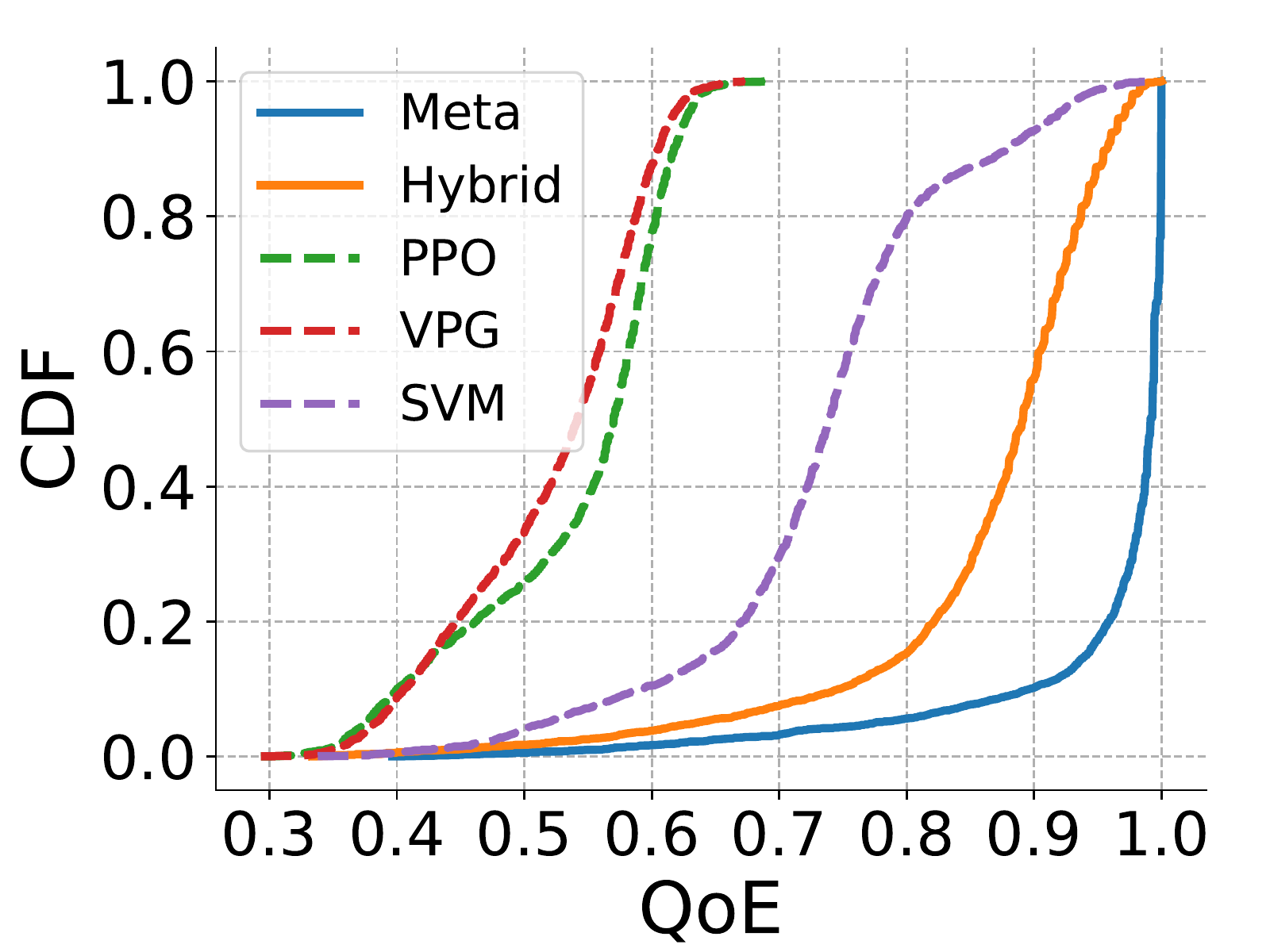}
		\caption{$(\eta_1, \eta_2) = (0.25, 1.0)$}
	\end{subfigure}
	~
	\begin{subfigure}[b]{0.22\linewidth}
		\centering
		\includegraphics[width=1.1\linewidth]{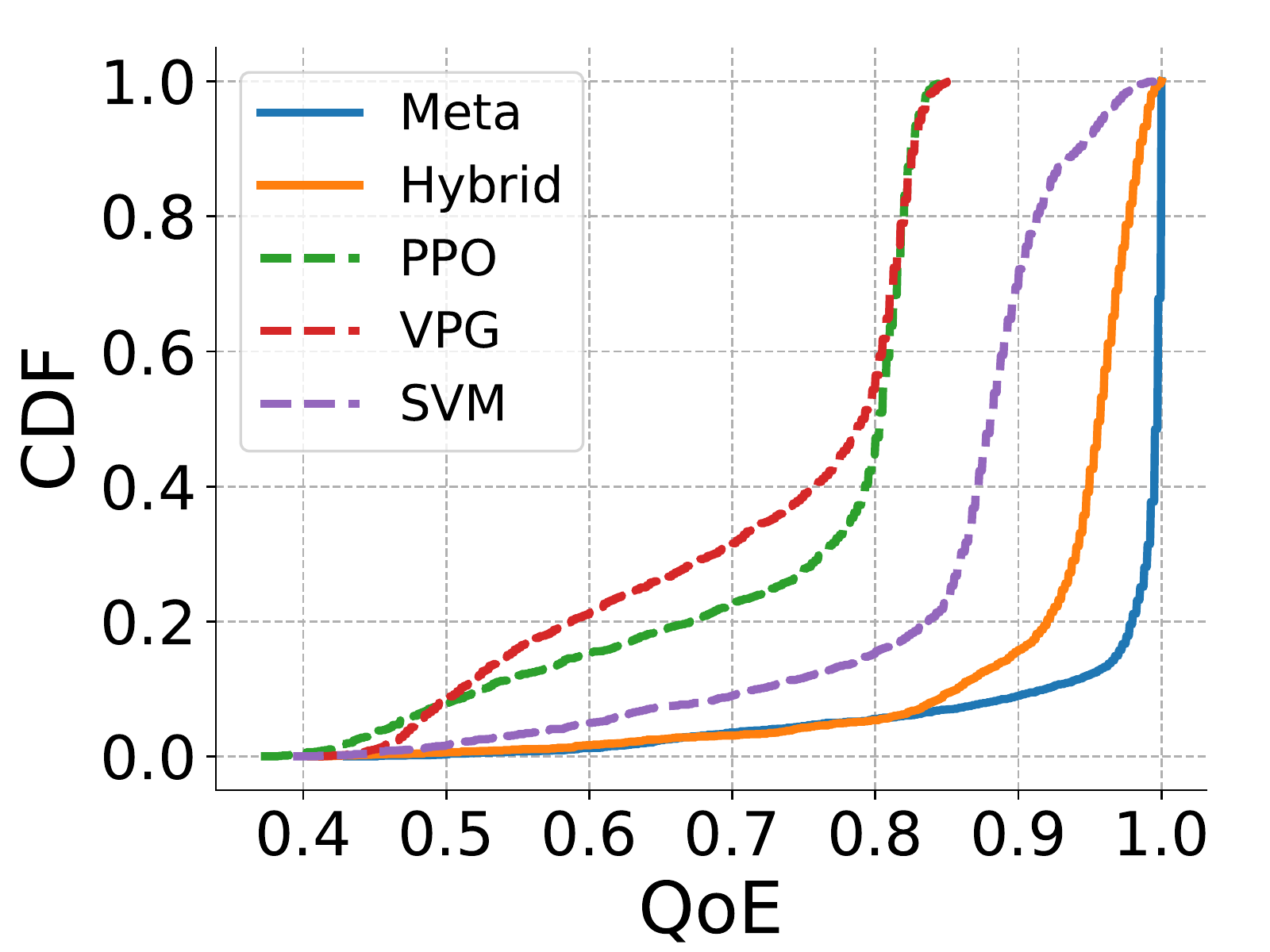}
		\caption{$(\eta_1, \eta_2) = (1.0, 1.0)$}
	\end{subfigure}
	~
	\begin{subfigure}[b]{0.22\linewidth}
		\centering
		\includegraphics[width=1.1\linewidth]{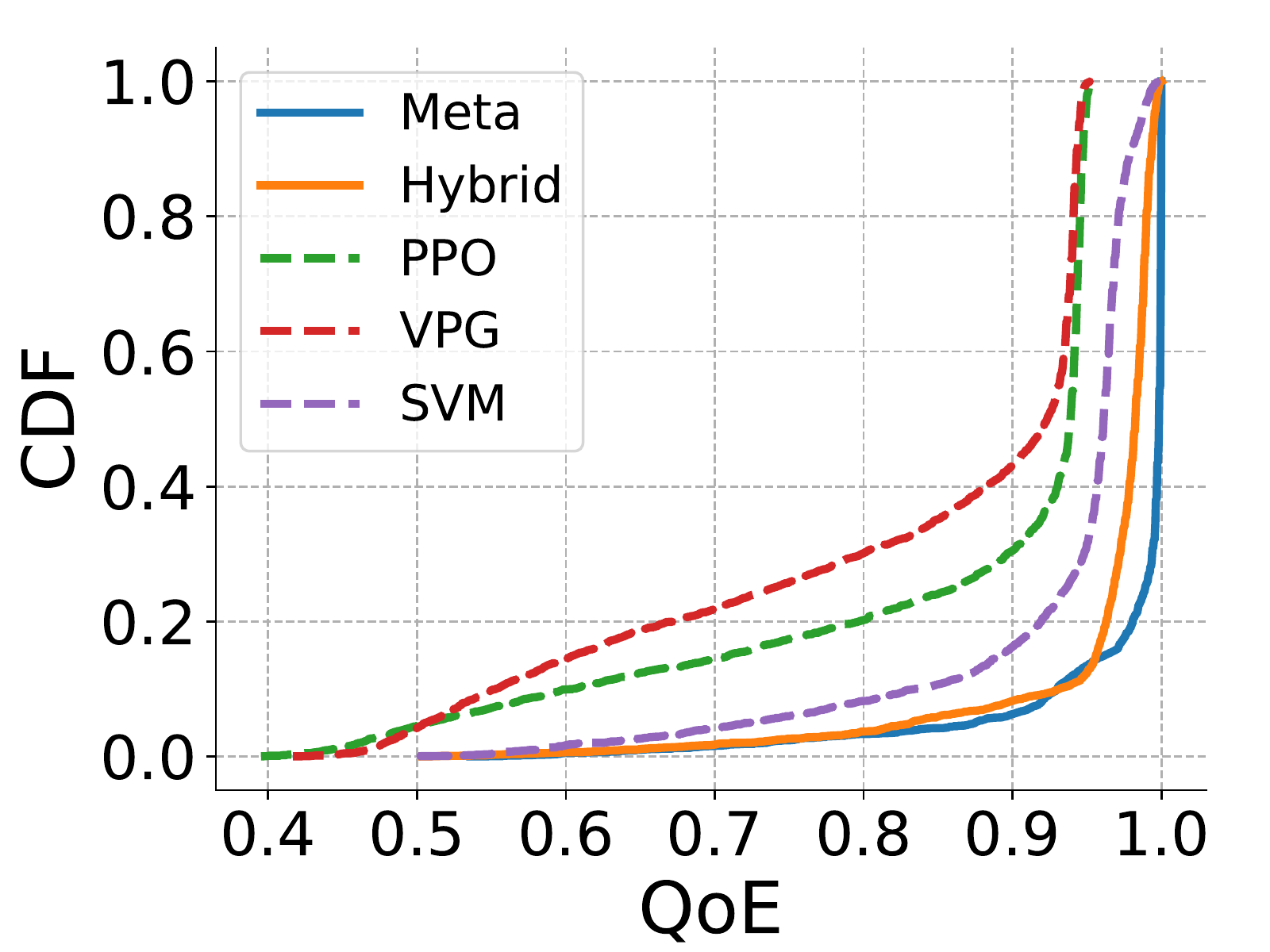}
		\caption{$(\eta_1, \eta_2) = (4.0, 1.0)$}
	\end{subfigure}
	\caption{CDF values of user QoE in different scenarios to show the trade-off between the system's latency and classification accuracy.}
	\label{fig:eta-varies}
\end{figure*}

We first illustrate the training performance of the proposed algorithm and the baselines in Fig.~\ref{fig:training-results}. In Fig.~\ref{fig:training-results}(a), we can observe the increase in QoE values of all the algorithms during 3,000 training episodes. 
The Meta-leaner can obtain the optimal QoE after only around 1,200 episodes while the Hybrid learner and SVM learner converge to lower QoE values at approximately $0.96$ and $0.91$, respectively, with longer convergence times. These converged QoE values for the PPO learner and VPG learner are approximately $0.81$. The gaps between the obtained QoE values of the learning algorithms can be further explained by Figs.~\ref{fig:training-results}(b) and \ref{fig:training-results}(c).

As observed in Fig.~\ref{fig:training-results}(b), the Meta-learner can obtain the $100\%$ classification accuracy while the Hybrid learner and SVM learner can only obtain around $82\%$ and $65\%$ classification accuracy, respectively. Notably, the PPO and VPG learners fail to classify the received BCI signals into appropriate classes, resulting in classification accuracy maintained at the chance level, i.e., $25\%$. The reason for the highly accurate classification results of the Meta learner is that it updates the meta-gradients based on (\ref{eq:meta-gradient-expand}) and (\ref{eq:meta-update}), which is equivalent to learning a distilled policy of $K$ separate Hybrid learners. The cost for this step is additional computations on $w$ mini-batches. However, in our experiments, we found that with a small value $w=3$, the Meta-learner achieves significant convergence speed, in exchange for a small computation overhead.
Note that the results in Fig.~\ref{fig:training-results}(b) are evaluated on the training set where the algorithms can obtain $100\%$ classification accuracy by training on the same training set multiple times. However, at the testing phase, we only use the trained models of the algorithms, which are obtained after 3,000 training episodes, to predict the test set of the BCI data on a single episode. As a result, the classification accuracy obtained on the test set, i.e., Figs \ref{fig:power-varies}, \ref{fig:cpu-varies}, and \ref{fig:subject-varies}, might be lower than the results obtained in Fig.~\ref{fig:training-results}.
 
The SVM baseline shows its superior performance over the PPO and VPG baselines thanks to its robust classification ability.
Another observation on the performance of the PPO and VPG baselines shows that they are not effective in dealing with the classification problem with only reinforcement learning design. With the number of class labels being $C=4$, the prediction accuracy of the PPO and VPG baselines are just slightly higher than the random prediction, i.e., $25\%$. Although the maximization of the QoE with reinforcement learning algorithms like PPO and VPG may result in reducing the classification loss theoretically, the exploration of such algorithms over the observation space makes the deep neural networks not have enough guidance to learn a good policy. This problem is known as a sparse reward problem in popular reinforcement learning settings \cite{sutton2018reinforcement}.
In Fig.~\ref{fig:training-results}(c), we can observe that all algorithms can converge to round-trip delay values that are less than the requirement $D_{\max} = 0.01$ second (i.e., 10 milliseconds). Thanks to the reinforcement learning techniques, all algorithms can learn the dynamics of radio and computing resources of the system. 

\subsubsection{Impacts of the QoE's weighting factors}
It is noted that the above training results are obtained with the setting $(\eta_1, \eta_2) = (1, 1)$. In other words, the importance of BCI classification and radio/computing resource allocation are similar. 
To evaluate the trade-off between the system's delay (which is directly affected by the resource allocation) and the classification accuracy, we consider different scenarios by changing the relative importance between the BCI classification and radio/computing resource allocation in Fig.~\ref{fig:eta-varies}.  The Cumulative Distribution Function (CDF) curves are obtained by averaging over 3,000 training episodes. In Fig.~\ref{fig:eta-varies}(a), we can observe that when the BCI classification is more important, i.e., $\eta_2 > \eta_1$, the QoE of the proposed algorithms is higher than those of the PPO and VPG baselines. For example, with the same CDF value at $0.6$, the proposed Meta-learner and Hybrid learner can achieve higher QoE values, i.e.,  $0.98$ and $0.91$, respectively. The results for the SVM, PPO, and VPG learners are $0.76$, $0.58$, and $0.55$, respectively.

In Fig.~\ref{fig:eta-varies}(b), when the values of $\eta_1$ and $\eta_2$ are similar, the proposed Meta-learner and Hybrid learner clearly outperform the baselines and achieve similar performance as the SVM baseline.
In addition, when $\eta_1$ increases, i.e., $\eta_1 > \eta_2$ in Fig.~\ref{fig:eta-varies}(c), the CDF curves of the PPO and VPG algorithms shift toward the right and become closer to the CDF curves of the Meta-learner, Hybrid learner, and the SVM baseline. The reason is that when the importance of the radio/computing resource allocation increases, the QoE values are less affected by the classification accuracy and the system's delay contributes more to the QoE.
From the three different settings above, we can conclude that our proposed algorithms are more robust, compared with the baseline algorithms.
Thanks to the ability to collectively store the training data, the SVM baseline can also achieve robust performance.  
In the rest of the simulations, we consider different settings with the values $(\eta_1, \eta_2) = (1, 1)$.

\subsubsection{Impacts of transmission power}
\begin{figure}[t]
	\centering
	\begin{subfigure}[b]{0.5\linewidth}
		\centering
		\includegraphics[width=1.0\linewidth]{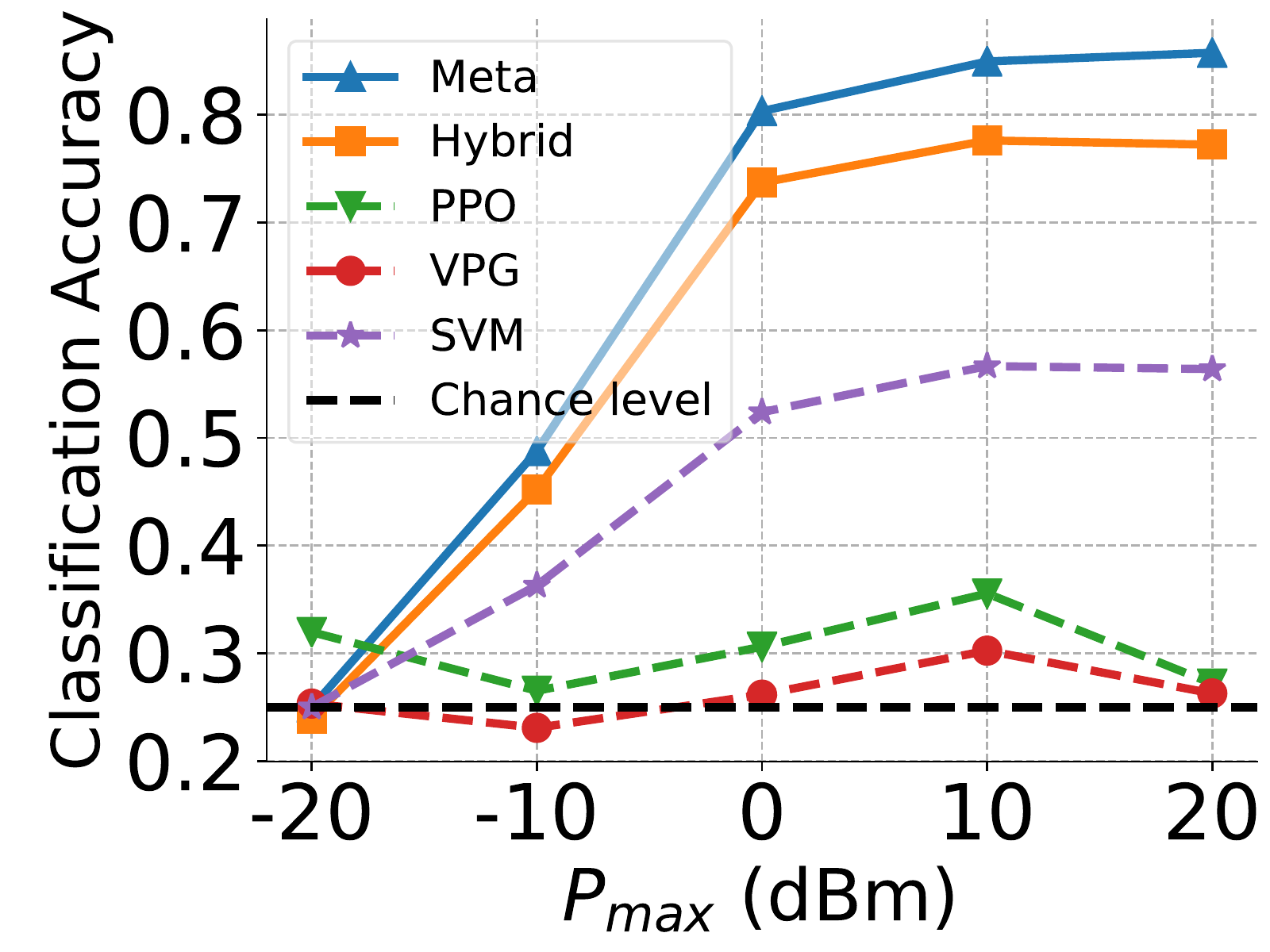}
		\caption{}
	\end{subfigure}%
	\begin{subfigure}[b]{0.5\linewidth}
		\centering
		\includegraphics[width=1.0\linewidth]{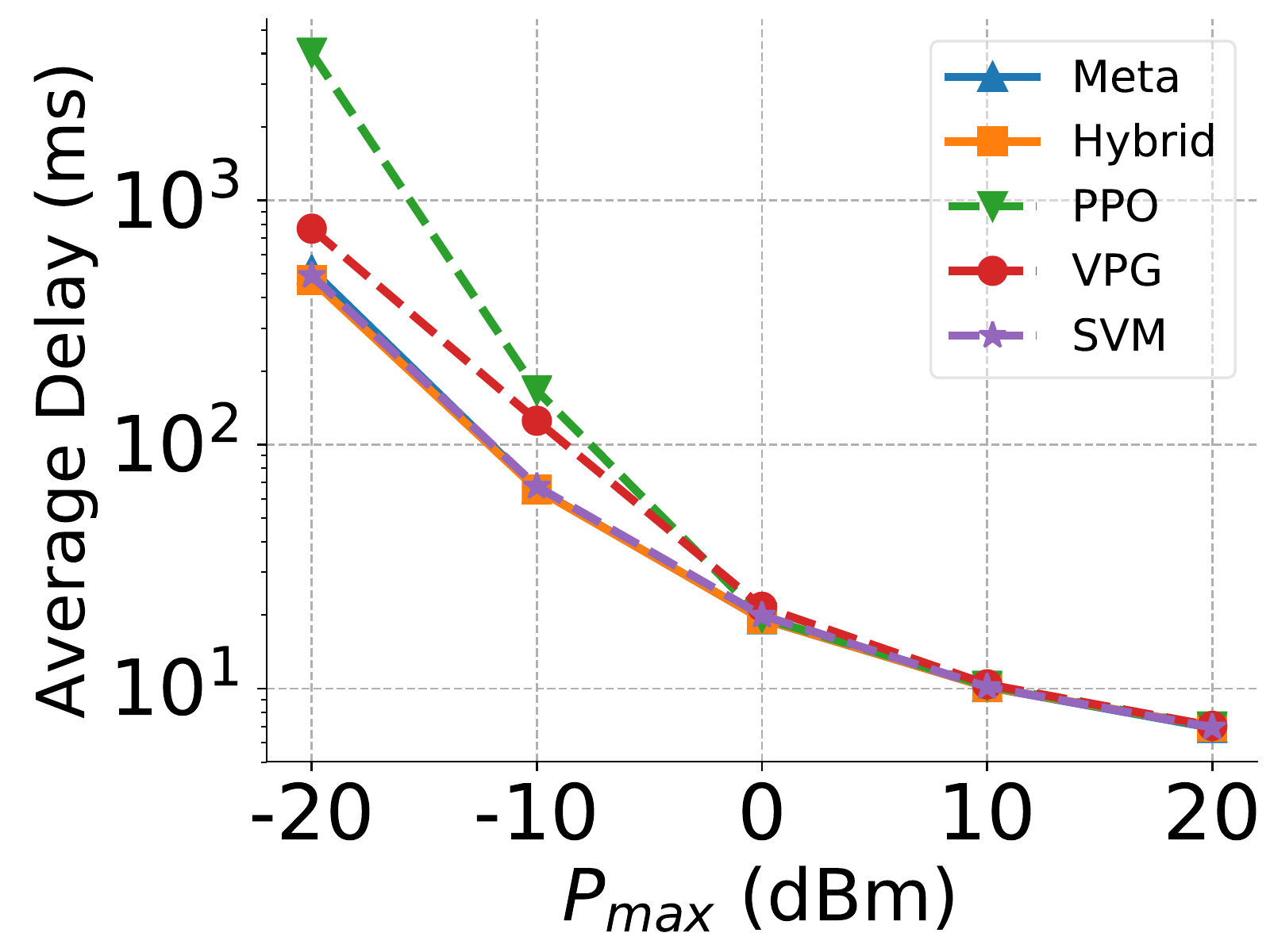}
		\caption{}
	\end{subfigure}
	\caption{(a) BCI classification accuracy and (b) round-trip VR delay of the algorithms with testing data when the maximum power of the headsets varies.} 
	\label{fig:power-varies}
\end{figure}

Next, we vary the maximum power value at the headsets of users, i.e., $P_{\max}$, to evaluate the impacts of the power allocation on the system performance. In Fig.~\ref{fig:power-varies}(a), we can observe that when the maximum power of the headsets increases, the accuracy of the prediction increases. The reason is that with the low level of power allocated to the radio resource blocks, the SINR at the WES may significantly decrease, resulting in the high packet error rate $\epsilon_k$ in (\ref{eq:epsilon}). For example, at the low values of $P_{max}$, e.g., $-20$ dBm and $-10$ dBm, the classification accuracy of the proposed Meta-learner and Hybrid learner are significantly decreased to approximate $25\%$ and $48\%$, respectively.
Nevertheless, our proposed Meta-learned achieves the best classification accuracy under most of the considered settings. 
Similar to the observation from Fig.~\ref{fig:training-results}(b), the classification accuracy values obtained by the SVM baseline are higher than those of the PPO and VPG baselines.

In Fig.~\ref{fig:power-varies}(b), we can observe that the increase of power results in the decrease of the round-trip VR delay. The latency values obtained by our proposed algorithms are the lowest among those of the baseline algorithms due to two main reasons. 
First, they utilize state-of-the-art actor-critic architecture and policy clipping techniques of PPO. Second, our new design in forwarding the losses through the actor network, critic network, and convolutional network, as described in Section~\ref{sec:learning-algorithms}, enables the realization of Hybrid learner and Meta-learner that the compound objective $Q_k$ involves distinct learning goals, i.e., BCI classification and radio, computing resource allocation; and thus facilitating the training process.

\subsubsection{Impacts of the computing capacity}
\begin{figure}[t]
	\centering
	\begin{subfigure}[b]{0.5\linewidth}
		\centering
		\includegraphics[width=1.0\linewidth]{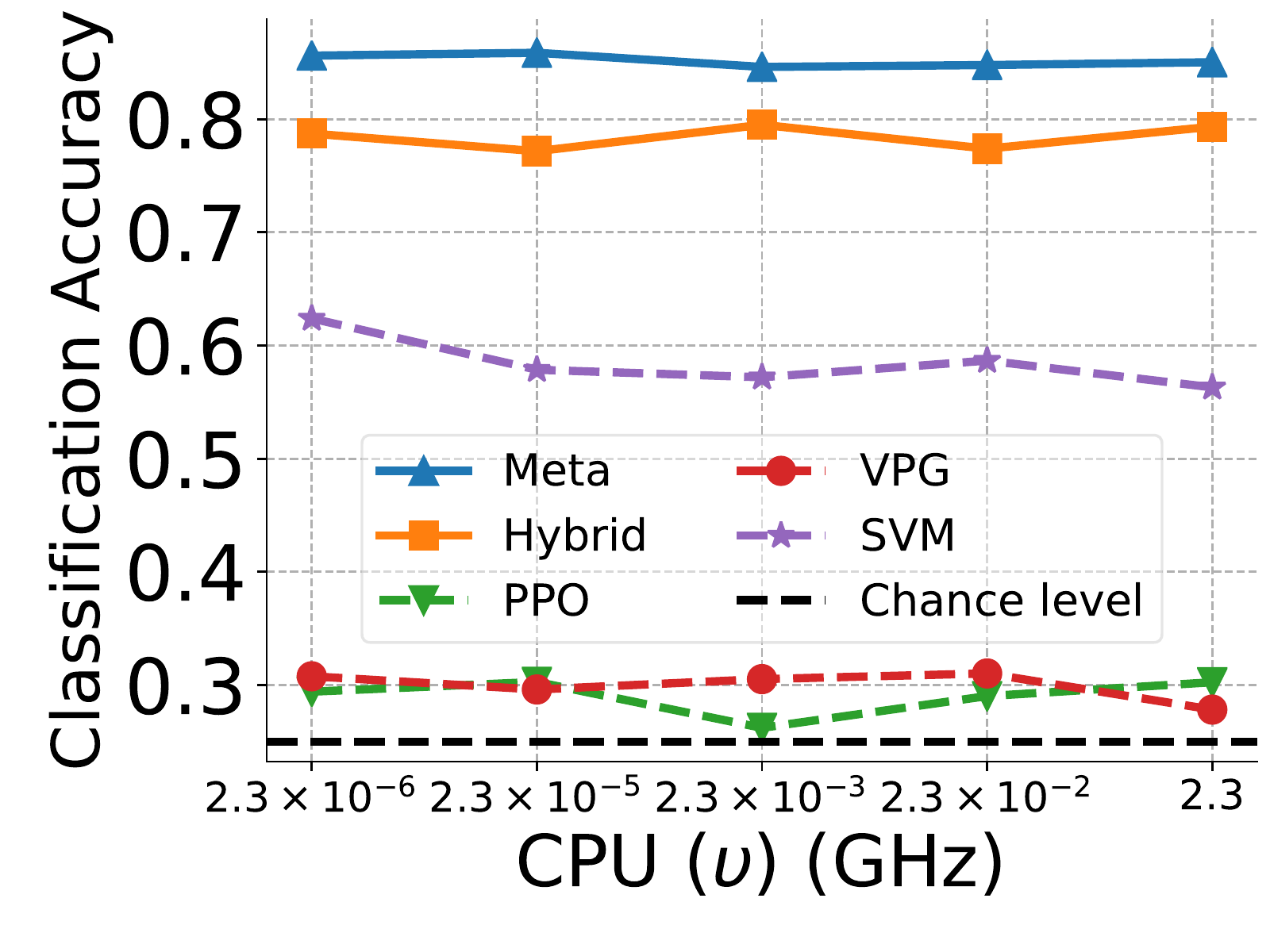}
		\caption{}
	\end{subfigure}%
	\begin{subfigure}[b]{0.5\linewidth}
		\centering
		\includegraphics[width=1.0\linewidth]{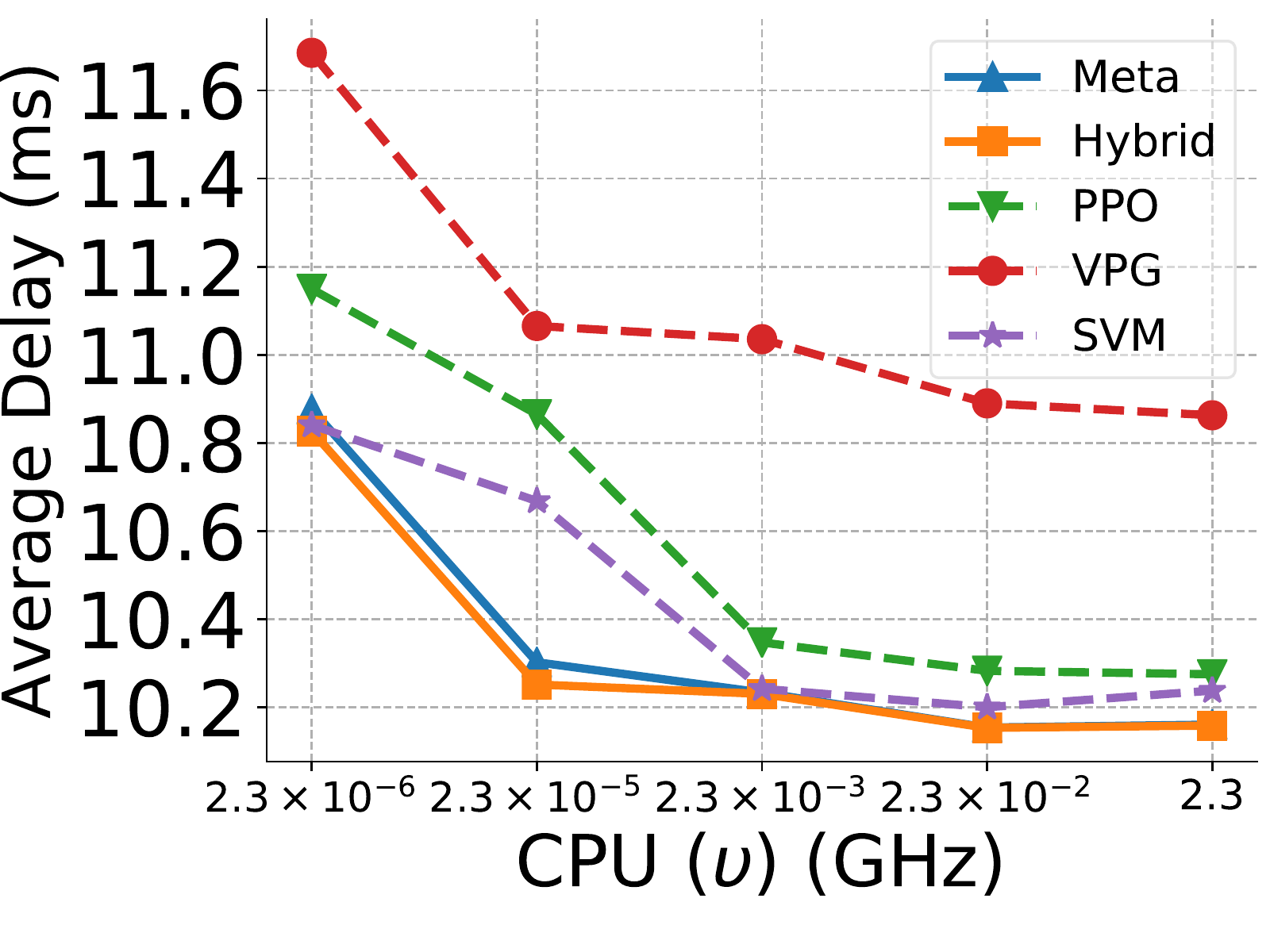}
		\caption{}
	\end{subfigure}
	\caption{(a) Classification accuracy and (b) round-trip VR delay of the algorithms with testing data when the CPU capacity of the WES varies.} 
		\label{fig:cpu-varies}
\end{figure}

Furthermore, we evaluate the impacts of the computing capacity of the WES on the system performance by increasing the CPU capacity of the WES from $2.3 \times 10^{-6}$ GHz to $2.3$ GHz.
In Fig.~\ref{fig:cpu-varies}(a), it can be observed that the increase in the CPU capacity of the WES does not have an impact on the classification accuracy. These results imply that with the limited CPU capacity, our proposed algorithms with deep neural networks still achieve good predictions on the BCI signals. Unlike our proposed algorithms with advanced architecture designs, the PPO and VPG baselines only obtain around $30\%$ of classification accuracy which is slightly higher than the chance level $25\%$. In Fig.~\ref{fig:cpu-varies}(b), we can observe that the increase in CPU capacity results in a decrease in the round-trip VR delay. The reason is that with lower CPU capacity, the WES takes a longer time to pre-precess VR contents for the users, thus yielding higher latency.
Note that the decrease in round-trip VR delay in Fig.~\ref{fig:cpu-varies}(b) is less significant than the decrease in Fig.~\ref{fig:power-varies}(b). The reason is that the transmit power has more impact on the system's delay, compared with the computing capacity.

\subsubsection{Impacts of the neurodiversity}
\begin{figure}[t]
	\centering
	\begin{subfigure}[b]{0.5\linewidth}
		\centering
		\includegraphics[width=1.0\linewidth]{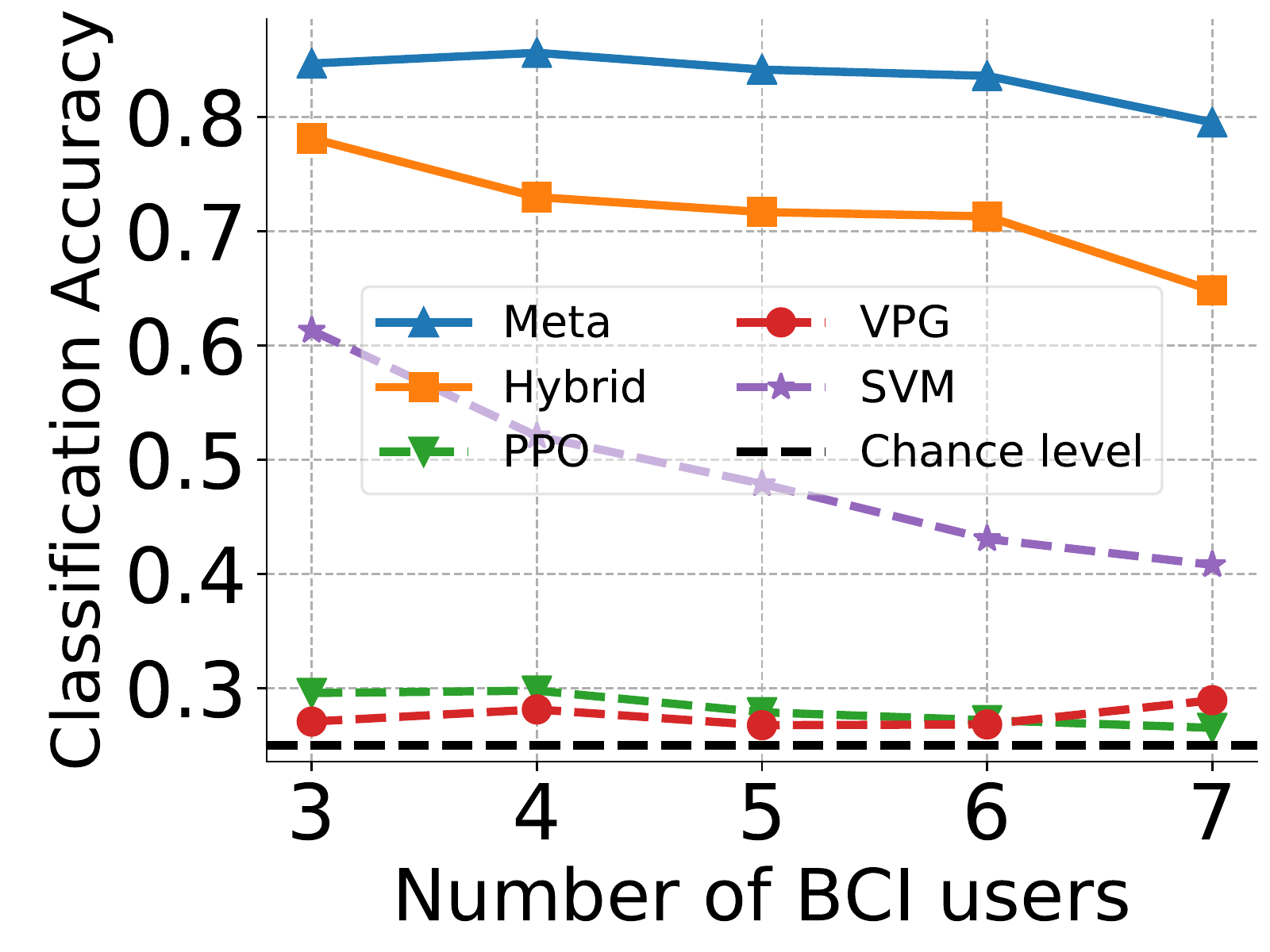}
		\caption{}
	\end{subfigure}%
	\begin{subfigure}[b]{0.5\linewidth}
		\centering
		\includegraphics[width=1.0\linewidth]{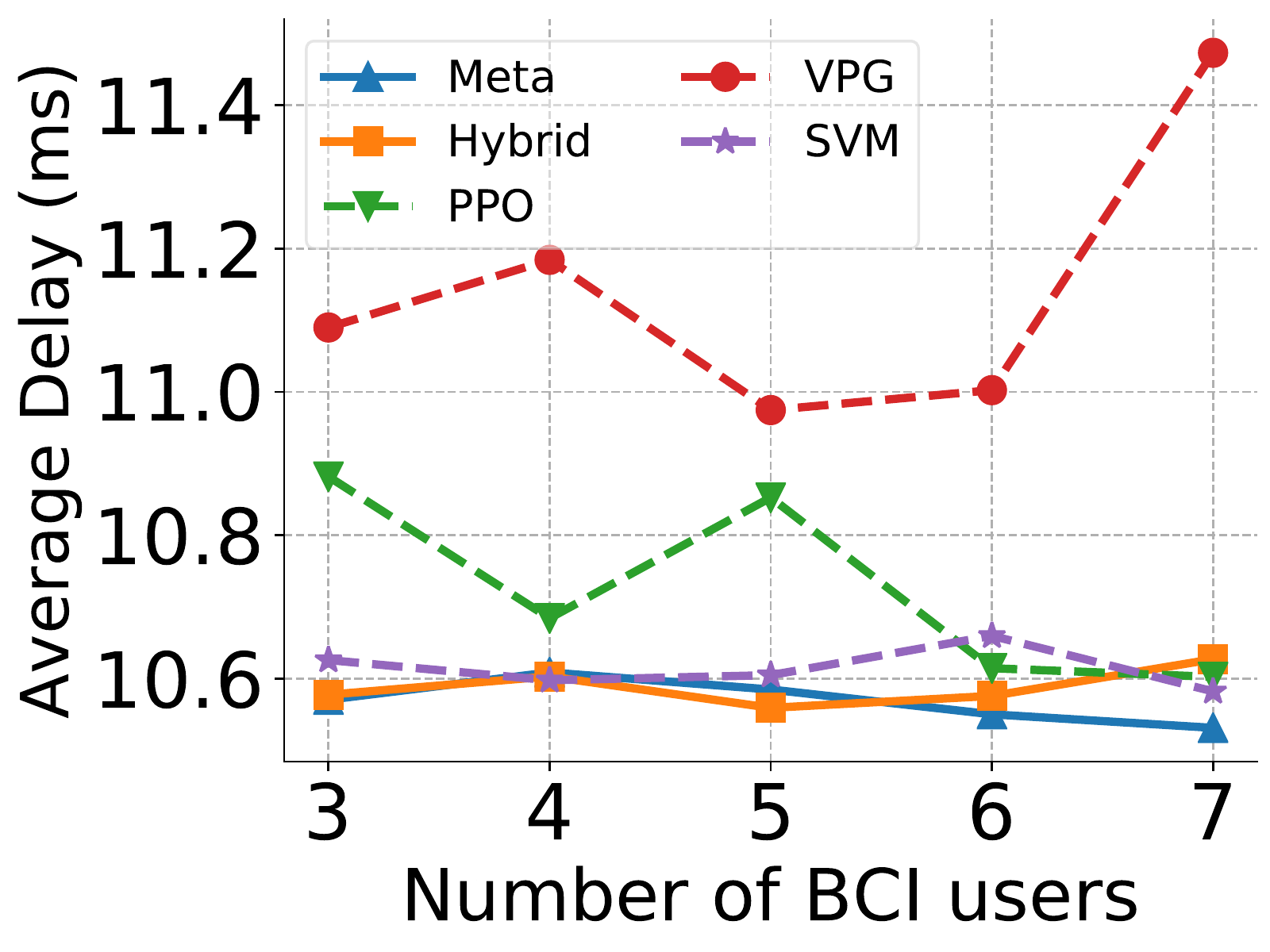}
		\caption{}
	\end{subfigure}
	\caption{(a) Classification accuracy and (b) round-trip VR delay of the algorithms with testing data when the number of BCI users varies.} 
	\label{fig:subject-varies}
\end{figure}

Finally, we evaluate the classification challenges caused by the diversity of BCI signals. In Fig.~\ref{fig:subject-varies}, we increase the number of BCI users from three to seven. In Fig.~\ref{fig:subject-varies}(a), the classification accuracy of the proposed Hybrid learner significantly decreases when the number of BCI users increases. 
The reason for this is that although the proposed Hybrid leaner with advanced architecture can obtain good prediction accuracy, the learner cannot deal with the diversity of the BCI signals as input data. 
As explained in Fig.~\ref{fig:eeg-example}, the BCI signals are highly individual so that the responses of different users on the same experimental instruction, e.g., fist movement, are different in both amplitude and phase. Therefore, a conventional convolutional neural network is not sufficient to learn from the neurodiversity of the BCI signals. 
Similarly, the classification accuracy values obtained by the baseline SVM are significantly decreased as the number of users increases. Even being trained with a larger memory of iteratively collected BCI data, the SVM baseline cannot achieve good classification accuracy. This observation shows the advantages of the deep neural network, i.e., CNN, over the conventional machine learning classifier as the SVM.

Unlike the Hybrid learner and SVM baseline, the proposed Meta-learner shows its capability of learning a distilled model that can better classify BCI signals from different users, reflected by the slightly decreasing classification accuracy. 
Specifically, with seven BCI users, our Meta-learner can achieve classification accuracy up to $78\%$.
The results suggest that the proposed Meta-learner is sample-efficient and practical in systems with limited storage capacity. 
Note that in the multi-person BCI classification setting above, our Meta-learner does not require any additional feature extraction methods to achieve similar accuracy results reported in \cite{kang2014bayesian, vezard2015eeg, zhang2017multi}.
Moreover, the works in \cite{kang2014bayesian, vezard2015eeg, zhang2017multi} only consider classification problem for the raw and noise-free BCI signals while we also consider the possible packet errors of transmitting BCI signals over noisy channels.
In Fig.~\ref{fig:subject-varies}(b), we can observe that the round-trip VR delay is not affected by the increase of BCI users. Our proposed algorithms also achieve lower latency, compared with the baseline algorithms. We observe that the fluctuations of the results over a small interval of latency could happen due to the inconsistency of the deep neural networks' optimizers. A promising solution to address this issue is to use a small number of hold-out data samples for additional validation and fine-tuning steps.  

\section{Conclusion}
\label{sec:conclusion}
In this work, we have introduced a novel over-the-air BCI-enabled framework for human-centric Metaverse applications.
The Digital Avatars can learn from human brain signals to predict the actions of the users under controlled permissions.
In addition, the novel system design enables the WES to jointly optimize the radio and computing resources of the system as well as the classification performance under the dynamics of the wireless environment and Metaverse users' behaviors. This was realized by the proposed hybrid learning algorithm and meta-learning algorithm to simultaneously address the mixed decision-making and classification problems. The hybrid learning algorithm has shown its effectiveness in handling mixed decision-making and classification problems of our system. This was thanks to the novel architecture which consists of three deep neural networks to split, compute, and backpropagate the losses. We have further proposed the meta-learning algorithm as an improved version of the hybrid learning algorithm to deal with the neurodiversity of the brain signals from different users.
Extensive experiments with real-world datasets showed that our proposed algorithms can achieve prediction accuracy up to $84\%$. More interestingly, our proposed algorithms also reduced the VR latency of the system, resulting in the potential reduction of VR sickness and enhancing user QoE in future Metaverse applications. Several potential research directions could be extended from our framework. For example, queueing models or forward error correcting codes can be utilized for eliminating serve lossy conditions of the channel model, e.g.,  deep fading, thus achieving high-reliability communications of BCI signals.

\appendices 
\section{Details of the EEG Dataset}
\label{sec:appendix}
\begin{figure}
\centering
\includegraphics[width=1.0\linewidth]{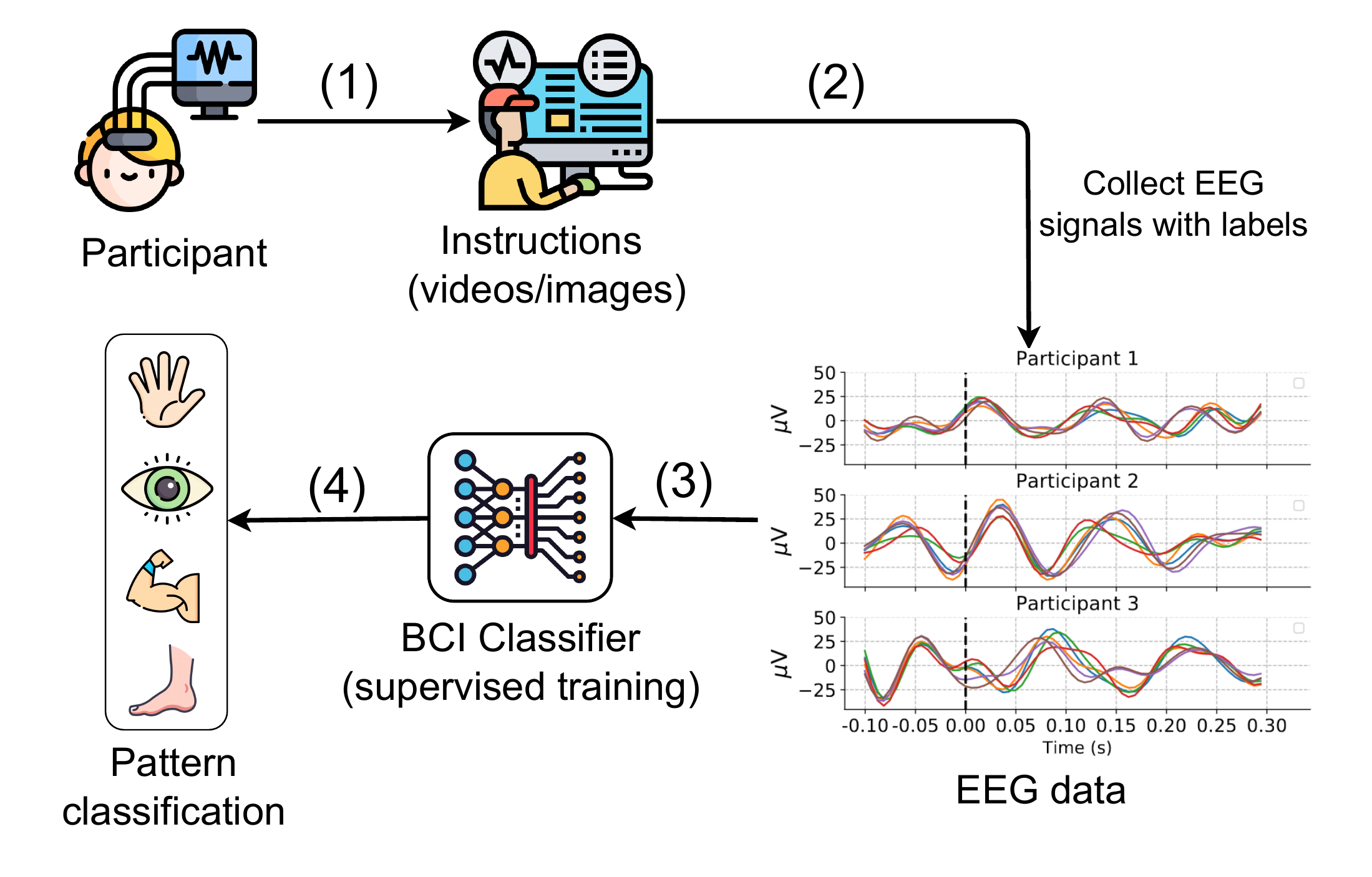}
\caption{The process of collecting EEG data in a motor imagery decoding experiment. The dataset in \cite{goldberger2000physiobank} is obtained by following the same process.}
\label{fig:motor-imagery}
\end{figure}

Overall, the dataset in \cite{goldberger2000physiobank} is collected by following an imagery decoding experiment. Motor imagery decoding in BCI research refers to the process of interpreting or decoding mental representations of motor actions from neural signals generated during the imagination of those actions. In other words, it involves extracting information from the brain related to imagined movements, such as the intention to move a limb, without any physical movement actually occurring.
The process of obtaining the EEG data from the motor imagery decoding experiments is illustrated in Fig.~\ref{fig:motor-imagery} and can be described as follows.

The first step (1) in Fig.~\ref{fig:motor-imagery} is to place EEG sensors on the participant's scalp. The positions of the EEG sensors follow some standards such as international 10-10 system or 10-20 system \footnotemark.
\footnotetext{\url{https://en.wikipedia.org/wiki/10\%E2\%80\%9320_system_(EEG)}}
Next, the participant is asked to imagine performing specific motor tasks, such as moving a hand, foot, or another body part, without actually executing the movement. The instructions are usually shown in the form of computer software with images/videos, which can be synchronized with the EEG headset to produce accurate measurements.

Next, in step (2) in Fig.~\ref{fig:motor-imagery}, the EEG headset records the brain activities of these mental actions. The raw EEG signals of all the EEG sensors, e.g., 64 sensors in \cite{goldberger2000physiobank}, are synchronized with labels. In addition, the EEG signals can also be divided into epochs, in which each epoch corresponds to a specific motor imagery action (or label), i.e., imagined opening fist, moving arms, etc. An example of the EEG signals in an epoch is illustrated in Fig.~\ref{fig:eeg-example}.

In step (3), the collected EEG signals are fed into a classifier, i.e., a decoding algorithm which is usually a machine learning or deep learning classifier, to translate these EEG signals/features into meaningful information about the intended motor actions. The output of the classifier will be further used for external tasks such as controlling a computer cursor, or a robotic arm, allowing individuals to interact with the environment based on their mental representations of movement, i.e., step (4).

The above steps can be repeatedly performed with multiple participants in multiple tasks. In particular, the dataset \cite{goldberger2000physiobank} contains EEG signals of 109 participants with 10 motor imagery tasks. The large number of participants can ensure unbiased performance for the classifier, while the number of imagery motor actions is sufficiently large for fundamental tasks.


\begin{IEEEbiography}[{\includegraphics[width=1in,height=1.25in,clip,keepaspectratio]{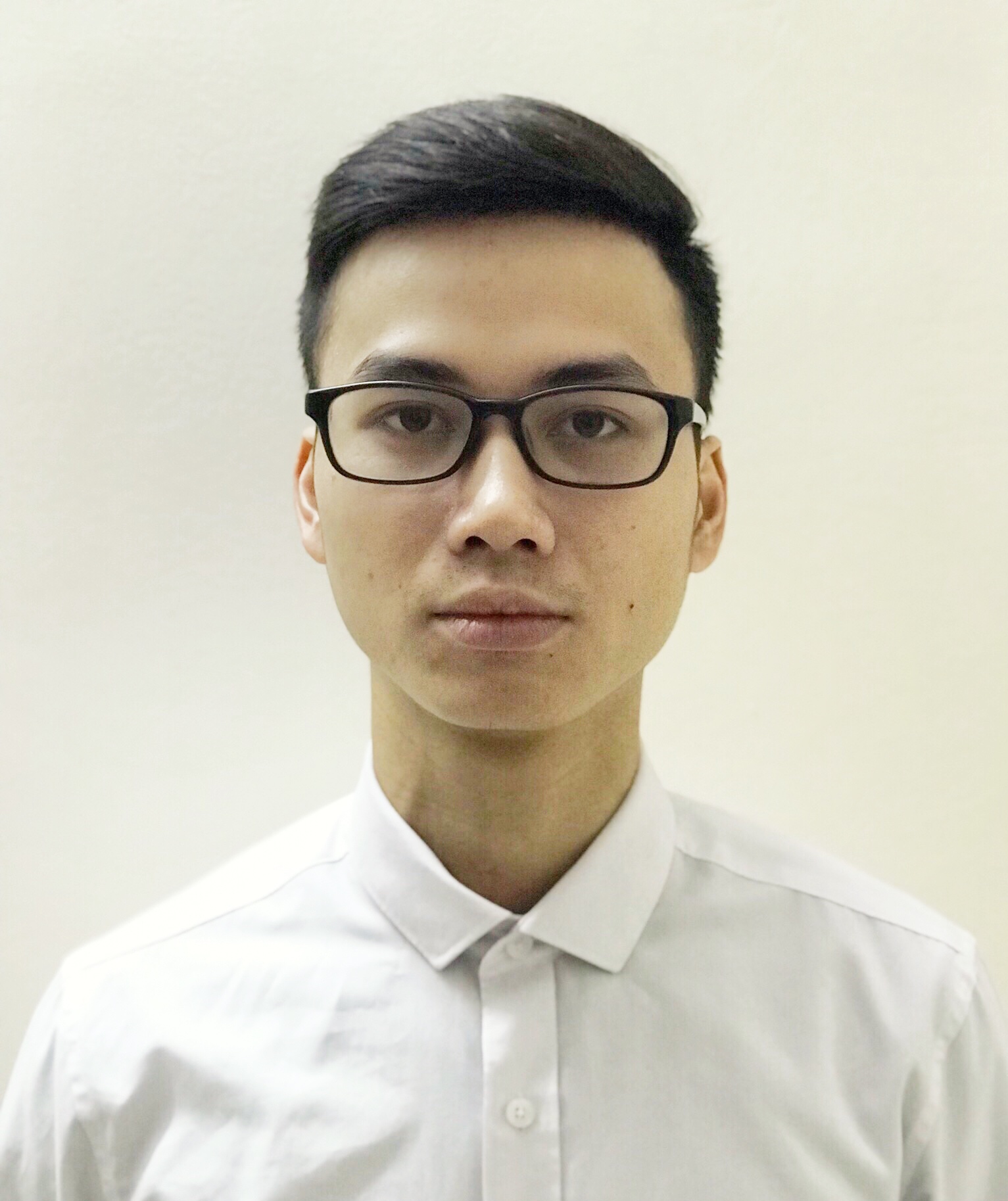}}]%
{Nguyen Quang Hieu} received the B.E. degree in Hanoi University of Science Technology, Vietnam in 2018. He is currently a Ph.D. student at School of Electrical and Data Engineering, University of Technology (UTS), Sydney, Australia. Before joining UTS, he was a research assistant at School of Computer Science and Engineering, Nanyang Technological University, Singapore. His research interest includes wireless communications and machine learning.
\end{IEEEbiography}

\vspace{-1.2cm}
\begin{IEEEbiography}[{\includegraphics[width=1in,height=1.25in,clip,keepaspectratio]{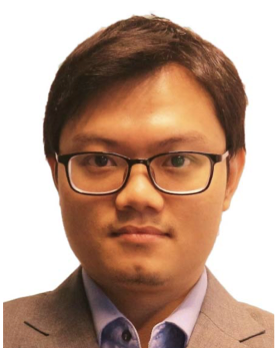}}]{Dinh Thai Hoang} (M'16, SM'22) is currently a faculty member at the School of Electrical and Data Engineering, University of Technology Sydney, Australia. He received his Ph.D. in Computer Science and Engineering from the Nanyang Technological University, Singapore 2016. His research interests include emerging wireless communications and networking topics, especially machine learning applications in networking, edge computing, and cybersecurity. He has received several precious awards, including the Australian Research Council Discovery Early Career Researcher Award, IEEE TCSC Award for Excellence in Scalable Computing for Contributions on “Intelligent Mobile Edge Computing Systems” (Early Career Researcher), IEEE Asia-Pacific Board (APB) Outstanding Paper Award 2022, and IEEE Communications Society Best Survey Paper Award 2023.  He is currently an Editor of IEEE TMC, IEEE TWC, IEEE TCCN, IEEE TVT, and IEEE COMST.
\end{IEEEbiography}

\vspace{-1.2cm}
\begin{IEEEbiography}[{\includegraphics[width=1in,height=1.25in,clip,keepaspectratio]{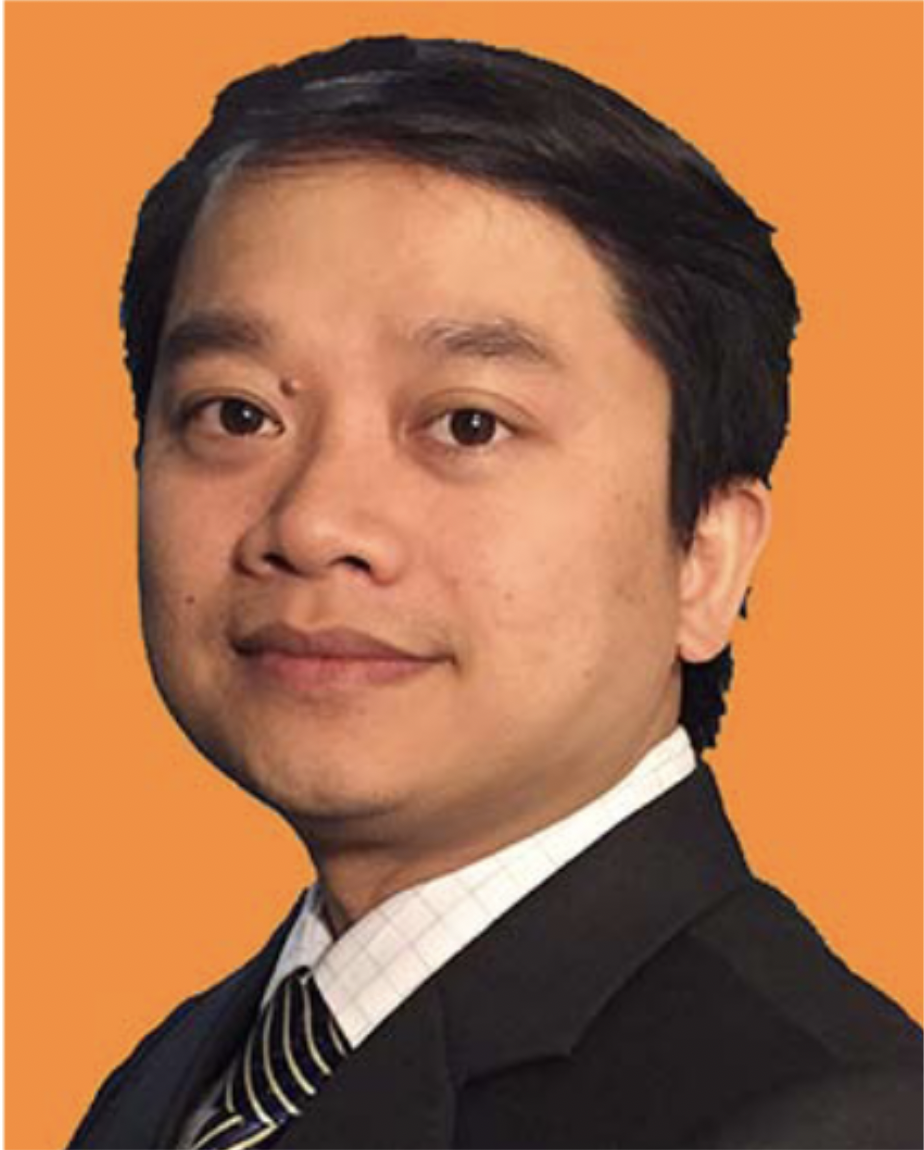}}]{Diep N. Nguyen}
 (Senior Member, IEEE) received the M.E. degree in electrical and computer engineering from the University of California at San Diego (UCSD), La Jolla, CA, USA, in 2008, and the Ph.D. degree in electrical and computer engineering from The University of Arizona (UA), Tucson, AZ, USA, in 2013. He is currently the Head of 5G/6G Wireless Communications and Networking Lab, Director of Agile Communications and Computing group, Faculty of Engineering and Information Technology, University of Technology Sydney (UTS), Sydney, NSW, Australia. Before joining UTS, he was a DECRA Research Fellow with Macquarie University, Macquarie Park, NSW, Australia, and a Member of the Technical Staff with Broadcom Corporation, CA, USA, and ARCON Corporation, Boston, MA, USA, and consulting the Federal Administration of Aviation, Washington, DC, USA, on turning detection of UAVs and aircraft, and the U.S. Air Force Research Laboratory, USA, on anti-jamming. His research interests include computer networking, wireless communications, and machine learning application, with emphasis on systems' performance and security/privacy. Dr. Nguyen received several awards from LG Electronics, UCSD, UA, the U.S. National Science Foundation, and the Australian Research Council. He has served on the Editorial Boards of the IEEE Transactions on Mobile Computing, IEEE Communications Surveys \& Tutorials (COMST), IEEE Open Journal of the Communications Society, and Scientific Reports (Nature's).\end{IEEEbiography}
 
\vspace{-1.2cm}
\begin{IEEEbiography}[{\includegraphics[width=1in,height=1.25in,clip,keepaspectratio]{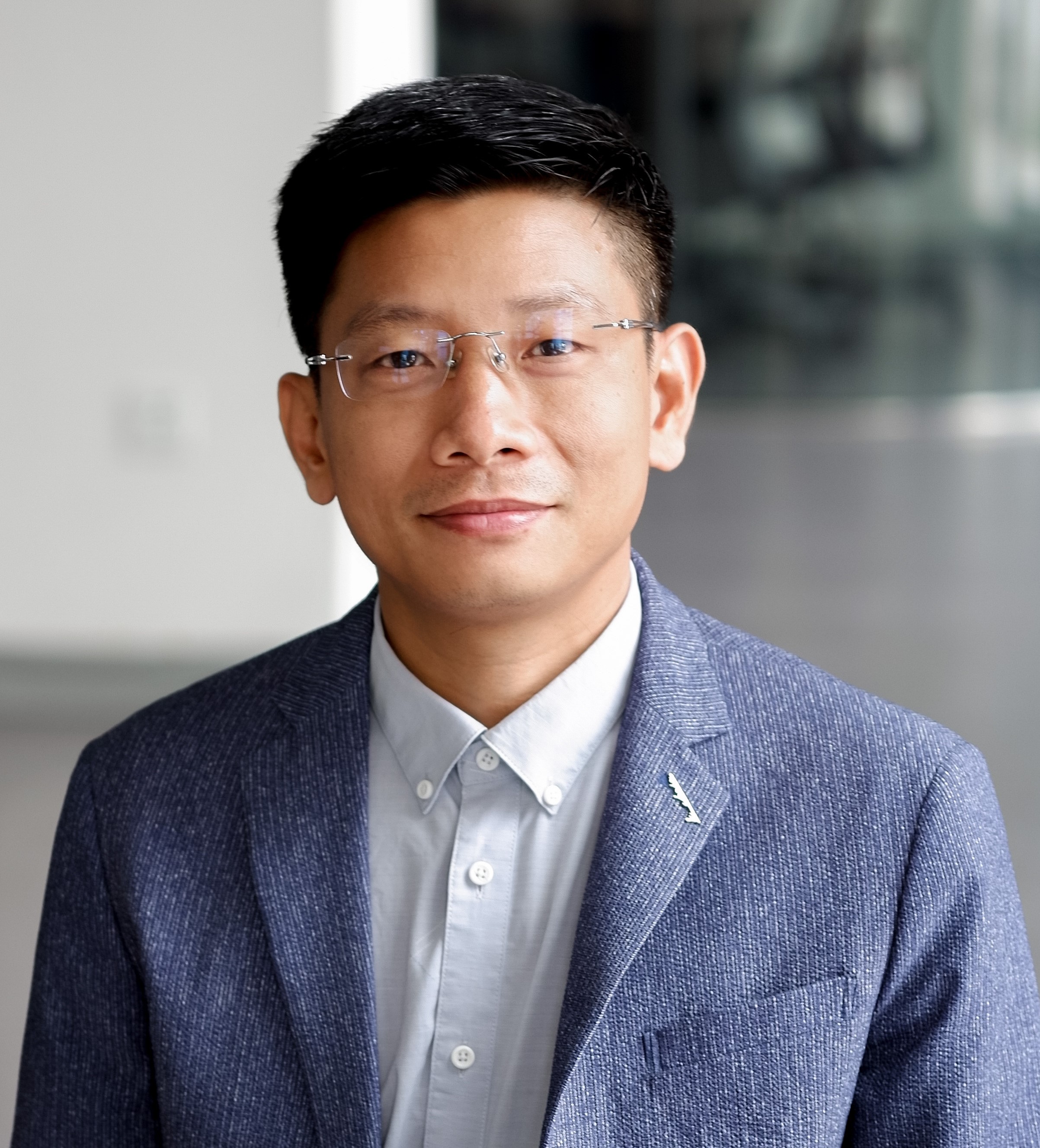}}]
{Van-Dinh Nguyen}(S’14-M’19-SM’23) received the B.E. degree in electrical engineering from Ho Chi Minh City University of Technology, Vietnam, in 2012 and the M.E. and Ph.D. degrees in electronic engineering from Soongsil University, Seoul, South Korea, in 2015 and 2018, respectively. Since 2022, he has been an Assistant Professor at VinUniversity, Vietnam. He was a Research Associate with the SnT, University of Luxembourg, a Postdoc Researcher and a Lecturer with Soongsil University, a Postdoctoral Visiting Scholar with University of Technology Sydney,  and a Ph.D. Visiting Scholar with Queen’s University Belfast, U.K. His current research activity is focused on the mathematical modeling of 5G/6G cellular networks, edge/fog computing, and AI/ML solutions for wireless communications. He is a Senior Editor for IEEE Communications Letters and an Associate Editor for IEEE Systems Journal and IEEE Open Journal of the Communications Society.
\end{IEEEbiography}

\begin{IEEEbiography}[{\includegraphics[width=1in,height=1.25in,clip,keepaspectratio]{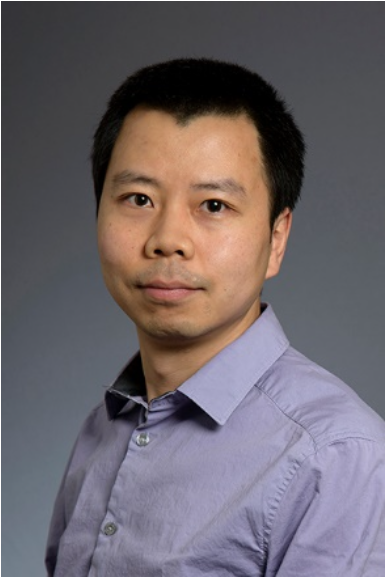}}]
{Yong Xiao}(Senior Member, IEEE) received his B.S. degree in electrical engineering from China University of Geosciences, Wuhan, China in 2002, M.Sc. degree in telecommunication from Hong Kong University of Science and Technology in 2006, and his Ph. D degree in electrical and electronic engineering from Nanyang Technological University, Singapore in 2012. He is now a professor in the School of Electronic Information and Communications at the Huazhong University of Science and Technology (HUST), Wuhan, China. He is also with Peng Cheng Laboratory, Shenzhen, China and Pazhou Laboratory (Huangpu), Guangzhou, China. He is the associate group leader of the network intelligence group of IMT-2030 (6G promoting group) and the vice director of 5G Verticals Innovation Laboratory at HUST. Before he joins HUST, he was a research assistant professor in the Department of Electrical and Computer Engineering at the University of Arizona where he was also the center manager of the Broadband Wireless Access and Applications Center (BWAC), an NSF Industry/University Cooperative Research Center (I/UCRC) led by the University of Arizona. His research interests include machine learning, game theory, distributed optimization, and their applications in semantic communications, semantic-aware networking, cloud/fog/mobile edge computing, green communication systems, and Internet-of-Things (IoT).
\end{IEEEbiography}

\vspace{-13.0cm}
\begin{IEEEbiography}[{\includegraphics[width=1in,height=1.25in,clip,keepaspectratio]{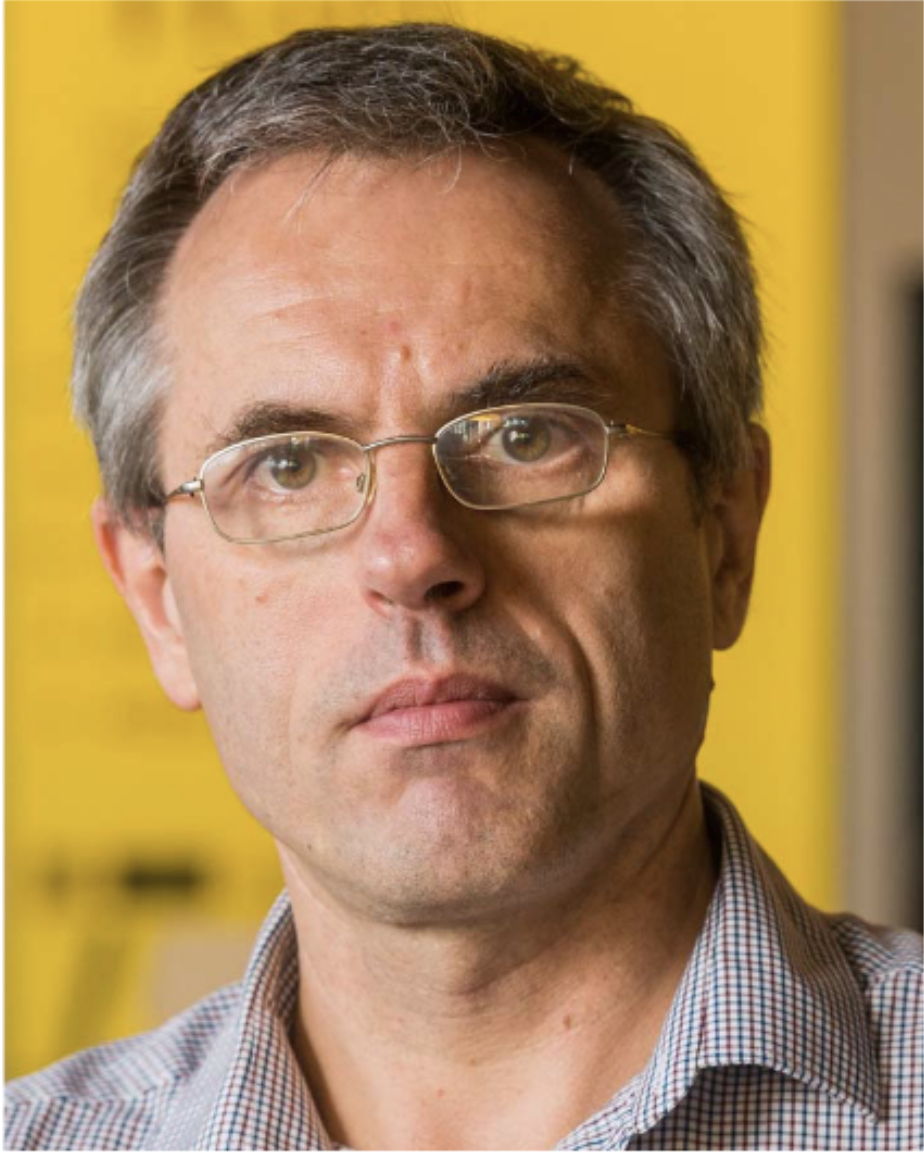}}]%
{Eryk Dutkiewicz} (Senior Member, IEEE) received his B.E. degree in Electrical and Electronic Engineering from the University of Adelaide in 1988, his M.Sc. degree in Applied Mathematics from the University of Adelaide in 1992 and his PhD in Telecommunications from the University of Wollongong in 1996. His industry experience includes management of the Wireless Research Laboratory at Motorola in early 2000's. Prof. Dutkiewicz is currently the Head of School of Electrical and Data Engineering at the University of Technology Sydney, Australia. He is a Senior Member of IEEE. He also holds a professorial appointment at Hokkaido University in Japan. His current research interests cover 5G/6G and IoT networks.
\end{IEEEbiography}
\end{document}